\documentclass[twoside,twocolumn,english,pra,show pacs]{revtex4}
\usepackage[T1]{fontenc}
\usepackage[latin9]{inputenc}
\usepackage{amstext}
\usepackage{amssymb}
\usepackage{graphicx}
\usepackage{esint}

\makeatletter
\@ifundefined{textcolor}{}
{%
 \definecolor{BLACK}{gray}{0}
 \definecolor{WHITE}{gray}{1}
 \definecolor{RED}{rgb}{1,0,0}
 \definecolor{GREEN}{rgb}{0,1,0}
 \definecolor{BLUE}{rgb}{0,0,1}
 \definecolor{CYAN}{cmyk}{1,0,0,0}
 \definecolor{MAGENTA}{cmyk}{0,1,0,0}
 \definecolor{YELLOW}{cmyk}{0,0,1,0}
}

\usepackage{multirow}

\makeatother

\usepackage{babel}
\begin{document}

\title{Effect of boson on-site repulsion on the superfluidity \\in
the boson-fermion-Hubbard model}

\author{A. S. Sajna\textsuperscript{} and R. Micnas\textsuperscript{}}
\begin{abstract}
We analyze the finite-temperature phase diagram of the boson-fermion-Hubbard
model with Feshbach converting interaction, using the coherent-state
path-integral method. We show that depending on the position of the bosonic band, this type of interaction, even
if weak, can drive the system into the resonant superfluid phase
in the strong bosonic interaction limit. It turns out that this phase
can exist for an arbitrary number of fermions (i.e., fermionic concentration
between 0 and 2) but with the bosonic particle number very close to an
integer value. We point out that the standard time-of-flight method
in optical lattice experiments can be an adequate technique to confirm
the existence of this resonant phase. Moreover, in the non-resonant
regime, the enhancement of the critical temperature of the superfluid phase
due to Feshbach interaction is also observed. We account for this interesting
phenomena for a hole- or particlelike pairing mechanism depending on
the system density and mutual location of the fermionic and bosonic bands. 
\end{abstract}

\pacs{67.85.Hj, 67.85.Bc, 64.70.Tg, 74.20.-z}

\address{Solid State Theory Division, Faculty of Physics, Adam Mickiewicz University, ulica Umultowska 85,
61-614 Pozna\'{n}, Poland}

\maketitle

\section{Introduction}

The boson-fermion-Hubbard model (BFHM) with resonant pairing mechanism
has a very long history in the context of high temperature superconductivity
(see, e.g. \cite{Ranninger1985,PhysRevB.36.180,RevModPhys.62.113,PhysRevB.40.6745,PhysRevB.42.4122,PhysRevB.55.3173,PhysRevB.64.104509,PhysRevB.63.134505,2002PhRvB..66j4516M,PhysRevB.67.134507,PhysRevB.70.184503,Mller2005,PhysRevB.73.214510,PhysRevB.83.214516,Micnas2014}
and references therein). Recently, the interest in this model has
been also extended to the ultracold atomic systems because they are
a versatile tool for simulating many-body physics \cite{Krutitsky_2016,RevModPhys.80.885,2012NatPh...8..267B}
and BFHM can be studied by using Feshbach resonance experiments in
which the BCS-BEC crossover is realized \cite{2012NatPh...8..267B,PhysRevLett.87.120406,CHEN2005,PhysRevLett.89.130402}.

The impact of strong bosonic interaction on the superfluid (SF) phase
in the lattice bosons system has been widely investigated in literature
in the terms of Bose-Hubbard model (BHM) (e.g. see \cite{Lewenstein2012}
and reference therein). However, the superfluidity in the regime of
strong bosonic repulsion in which Feshbach interaction is included
is much less understood. So far only hard-core limit \cite{PhysRevB.76.184507,PhysRevB.74.094511,PhysRevB.67.174521,RevModPhys.62.113,PhysRevB.36.180}
and some qualitative studies have been performed \cite{Zhou2006-BFM}.
Therefore in this paper, quantitative investigation of the non-zero
temperature BFHM phase diagram with finite bosonic repulsion interaction
is carried out, which is relevant for working out realistic experimental
conditions. The effective field theory description of the BFHM is
constructed by using the coherent state path integral formalism. This
analytical method seems to be a good starting point for analysis of
BFHM because it provides a reasonable description of the standard
Fermi Hubbard model at weak inter-particle interaction (i.e. in the
BCS regime) \cite{Altland:2010ww} and it also gives a correct description
of BHM \cite{2005PhRvA..71c3629S}. In this paper, we show that besides
the standard superfluid phase which is governed by the pure bosonic
correlation mechanism present in BHM, there appears also a resonant
superfluid (RSF) phase due to Feshbach resonance phenomena. Moreover,
we explain that the standard superfluid phase (not RSF) is enhanced
by the hole or particle pairing mechanism of fermions. The results
allow us to discuss experimental proposal for possible investigation
of RSF phase in BFHM.

In the following sections, we first describe the model and the coherent
state path integral method applied (Sec. II). Then, in Sec. III, we
use this method in analysis of the finite temperature phase diagram
of BFHM and its thermodynamic quantities. At the end of Sec. III we
also discuss experimental setups that could be used to prove some
results of our theory. Finally in Sec. IV we give a summary of our
work. Moreover, Appendix A and B contains additional investigation
of BFHM model within the operator approach.

\section{Model and method \label{sec:Model}}

\subsection{Model}

We consider the boson-fermion Hubbard model (BFHM) with converting
interaction energy $I$ whose Hamiltonian is given by \cite{PhysRevB.76.184507,Micnas2014}

\begin{eqnarray}
 &  & H=-\sum_{ij\sigma}\left(t_{ij}+\mu\delta_{ij}\right)c_{i\sigma}^{\dagger}c_{j\sigma}-V\sum_{i}c_{i\uparrow}^{\dagger}c_{i\downarrow}^{\dagger}c_{i\downarrow}c_{i\uparrow}\nonumber \\
 &  & -\sum_{ij}\left(J_{ij}+\mu^{*}\delta_{ij}\right)b_{i}^{\dagger}b_{j}+\frac{U}{2}\sum_{i}b_{i}^{\dagger}b_{i}^{\dagger}b_{i}b_{i}\nonumber \\
 &  & +I\sum_{i}\left[c_{i\uparrow}^{\dagger}c_{i\downarrow}^{\dagger}b_{i}+b_{i}^{\dagger}c_{i\downarrow}c_{i\uparrow}\right],\label{eq:BFM-hamiltonian}
\end{eqnarray}
where $\mu$ is the chemical potential, $\mu^{*}=2\mu-2\Delta_{B}$
and $\sigma$ is a spin$-\frac{1}{2}$ index ($\sigma\in\left\{ \uparrow,\downarrow\right\} $).
$c_{i\sigma}$ ($c_{i\sigma}^{\dagger}$) is fermionic annihilation
(creation) operator at site $i$ with spin $\sigma$ and $b_{i}$
($b_{i}^{\dagger}$) is bosonic annihilation (creation) operator at
site $i$. The hopping energies for fermions and bosons are $t_{ij}$
and $J_{ij}$, respectively. Throughout this work we restrict hopping
parameters to the nearest-neighbour sites. Moreover, $U$ denotes
the on-site interaction energy of bosons which will be treated exactly
during calculations and $V$ is the efermionic on-site interaction
strength. The bottom of bosonic band is shifted by $2\Delta_{B}$
parameter which could be tuned in ultracold atoms experiments with
the Feshbach resonance \cite{2012NatPh...8..267B,PhysRevLett.87.120406,PhysRevLett.89.130402,PhysRevA.67.033603}.

Interestingly, if we assume $I=0$ and independent chemical potentials, the BFHM Hamiltonian (Eq. (\ref{eq:BFM-hamiltonian})) describes two independent models  i.e. the fermionic and bosonic Hubbard models. However, in the presence of finite resonant interaction ($I\neq 0$), there is only one phase transition from the superfluid phase which we will show shortly.

Further, in the case of $U=V=0$ the model described by the Hamiltonian in
Eq. (\ref{eq:BFM-hamiltonian}) has been investigated earlier in the
continuum and lattice systems \cite{Micnas2014,CHEN2005,PhysRevB.76.184507,PhysRevA.67.063612,PhysRevB.53.R11961,PhysRevB.56.8330}.
Moreover, when $U\rightarrow\infty$ the hard-core bosonic limit is
obtained for which bosonic operators satisfy the Pauli spin $1/2$
commutations relations \cite{RevModPhys.62.113,PhysRevB.76.184507,PhysRevB.36.180,Micnas2014,2002PhRvB..66j4516M}.

In the coherent state path integral representation, the partition
function of BFHM reads

\begin{equation}
Z=\int\mathcal{D}\left[\bar{c},c,\bar{b,}b\right]e^{-\frac{1}{\hbar}S\left[\bar{c,}c,\bar{b,}b\right]},\label{eq:global-partition-function}
\end{equation}
where the action is given by

\begin{equation}
S\left[\bar{c,}c,\bar{b,}b\right]=S_{0}^{F}\left[\bar{c,}c\right]+S_{0}^{B}\left[\bar{b,}b\right]+S_{0}^{FB}\left[\bar{b,}b,\bar{c,}c\right]+S_{1}^{B}\left[\bar{b,}b\right].\label{eq:dzialanie_U_rowne_0-1}
\end{equation}
The denotation is related with perturbed and unperturbed parts of
the action which we exploit further, i.e. unperturbed parts are
\begin{eqnarray}
 &  & S_{0}^{F}\left[\bar{c,}c\right]=\int_{0}^{\hbar\beta}d\tau\left\{ \sum_{i\sigma}\bar{c}_{i\sigma}(\tau)\hbar\frac{\partial}{\partial\tau}c_{i\sigma}(\tau)\right.\nonumber \\
 &  & +\sum_{ij\sigma}\left(-t_{ij}-\mu\delta_{ij}\right)\bar{c}_{i\sigma}(\tau)c_{j\sigma}(\tau)\nonumber \\
 &  & \left.-V\sum_{i}\bar{c}_{i\uparrow}\left(\tau\right)\bar{c}_{i\downarrow}\left(\tau\right)c_{i\downarrow}\left(\tau\right)c_{i\uparrow}\left(\tau\right)\right\} ,
\end{eqnarray}
\begin{eqnarray}
 &  & S_{0}^{B}\left[\bar{b,}b\right]=\sum_{i}\int_{0}^{\hbar\beta}d\tau\left\{ \bar{b}_{i}(\tau)\hbar\frac{\partial}{\partial\tau}b_{i}(\tau)\right.\nonumber \\
 &  &\left. -\mu^{*}\bar{b}_{i}(\tau)b_{i}(\tau)+\frac{U}{2}\bar{b}_{i}(\tau)\bar{b}_{i}(\tau)b_{i}(\tau)b_{i}(\tau)\right\},
\end{eqnarray}
\begin{eqnarray}
S_{0}^{FB}\left[\bar{b,}b,\bar{c,}c\right] & = & I\sum_{i}\int_{0}^{\hbar\beta}d\tau\left[\bar{c}_{i\uparrow}(\tau)\bar{c}_{i\downarrow}(\tau)b_{i}(\tau)+\textrm{c.c.}\right],
\end{eqnarray}
and the part of the action which we will be treated approximately
is

\begin{eqnarray}
S_{1}^{B}\left[\bar{b,}b\right] & = & -\sum_{ij}\int_{0}^{\hbar\beta}d\tau J_{ij}\bar{b}_{i}(\tau)b_{j}(\tau)\,.\label{eq:S1B}
\end{eqnarray}
The fields $c_{i\sigma}\left(\tau\right)$, $\bar{c}_{i\sigma}\left(\tau\right)$
are Grassman variables, the $b_{i}\left(\tau\right)$, $\bar{b}_{i}\left(\tau\right)$
are complex variables, $\hbar$ is reduced Planck constant, $\beta=1/k_B T$ where $k_B$ and $T$ denote Boltzmann constant and temperature, respectively. Throughout this work we denote the complex
conjugation of arbitrary $x$ variable by $\bar{x}$.

\subsection{Effective action \label{sec: effective action}}

We are interested in the influence of the fermionic degrees of freedom
on the bosonic part in the BFHM model within the $J\ll U$ limit. 

In the first step, the therm describing the interaction between fermionic
particles is decoupled by the Hubbard-Stratonovich (HS) transformation
in the pairing channel which introduces $\Delta_{i}(\tau),\,\bar{\Delta}_{i}(\tau)$
fields \cite{Altland:2010ww}. Then $S_{0}^{F}\left[\bar{c,}c,\right]\rightarrow\tilde{S}_{0}^{F}\left[\bar{c,}c,\bar{\Delta},\Delta\right]$
where 
\begin{eqnarray}
 &  & \tilde{S}_{0}^{F}\left[\bar{c,}c,\bar{\Delta},\Delta\right]=\int_{0}^{\hbar\beta}d\tau\left\{ \sum_{i\sigma}\bar{c}_{i\sigma}(\tau)\hbar\frac{\partial}{\partial\tau}c_{i\sigma}(\tau)\right.\nonumber \\
 &  & -\sum_{i}\bar{c}_{i\uparrow}(\tau)\bar{c}_{i\downarrow}(\tau)\Delta_{i}(\tau)-\sum_{i}\bar{\Delta}_{i}(\tau)c_{i\downarrow}(\tau)c_{i\uparrow}(\tau)\nonumber \\
 &  & \left.+\sum_{ij\sigma}\left(-t_{ij}-\mu\delta_{ij}\right)\bar{c}_{i\sigma}(\tau)c_{j\sigma}(\tau)+\frac{1}{V}\sum_{i}\left|\Delta_{i}(\tau)\right|^{2}\right\} \,.
\end{eqnarray}
and for which the HS measure $\mathcal{D}\left[\bar{\Delta},\Delta\right]$
contains the determinant $\det\left[V^{-1}\right]$. Then, in the
$J\ll U$ limit, we decouple the term in the action from Eq. (\ref{eq:S1B})
which is proportional to $J$. It is performed by introducing the
HS transformation
\begin{eqnarray}
 &  & \sum_{ij}\int_{0}^{\hbar\beta}d\tau J_{ij}\bar{b}_{i}(\tau)b_{j}(\tau)\rightarrow-\sum_{ij}\int_{0}^{\hbar\beta}d\tau J_{ij}^{-1}\bar{\psi}_{i}(\tau)\psi_{j}(\tau)\nonumber \\
 &  & +\sum_{i}\int_{0}^{\hbar\beta}d\tau\bar{\psi}_{i}(\tau)b_{i}(\tau)+\sum_{i}\int_{0}^{\hbar\beta}d\tau\bar{b}_{i}(\tau)\psi_{i}(\tau)\,.\label{eq: 1 HS}
\end{eqnarray}
Going further, integrating out of bosonic fields $\bar{b}_{i}(\tau)$,
$b_{i}(\tau)$ is desirable. Before, we do that, we have to apply
some approximation of these fields since in the present form, the
action considered above, is non-integrable in $\bar{b}_{i}(\tau)$,
$b_{i}(\tau)$ because of the interaction term proportional to $U$.
Therefore we rewrite the partition function from Eq. (\ref{eq:global-partition-function})
to the following form
\begin{eqnarray}
 &  & Z=Z_{0}^{B}\det\left[\mathbf{J}^{-1}\right]\int\mathcal{D}\left[\bar{c},c,\bar{\psi},\psi,\bar{\Delta},\Delta\right]\nonumber \\
 &  & \times e^{-\frac{1}{\hbar}\sum_{ij}\int_{0}^{\hbar\beta}d\tau J_{ij}^{-1}\bar{\psi}_{i}(\tau)\psi_{j}(\tau)-\frac{1}{\hbar}\tilde{S}_{0}^{F}\left[\bar{c,}c,\bar{\Delta},\Delta\right]}\nonumber \\
 &  & \times\left\langle e^{-\frac{1}{\hbar}\sum_{i}\int_{0}^{\hbar\beta}d\tau\left(\left[-\bar{\psi}_{i}(\tau)+I\bar{c}_{i\uparrow}(\tau)\bar{c}_{i\downarrow}(\tau)\right]b_{i}(\tau)+c.c.\right)}\right\rangle _{0}^{B}
\end{eqnarray}
where $\mathbf{J}$ is the hopping matrix $J_{ij}$ which results
from the HS transformation in Eq. (\ref{eq: 1 HS}) and the statistical
average $\left\langle ...\right\rangle _{0}^{B}$ is defined as $\left(Z_{0}^{B}\right)^{-1}\int\mathcal{D}\left[\bar{b},b\right]...e^{-S_{0}^{B}\left[\bar{b,}b\right]/\hbar}$
with 
\begin{equation}
Z_{0}^{B}=\int\mathcal{D}\left[\bar{b},b\right]e^{-S_{0}^{B}\left[\bar{b,}b\right]/\hbar}.\label{eq:Z0B}
\end{equation}
Because $\psi_{i}\left(\tau\right)$, $\bar{\psi}_{i}\left(\tau\right)$
fields have quadratic form with linear terms we can make the shift
$\psi_{i}(\tau)\rightarrow\psi_{i}(\tau)+Ic_{i\downarrow}(\tau)c_{i\uparrow}(\tau)$
and $\bar{\psi}_{i}(\tau)\rightarrow\bar{\psi}_{i}(\tau)+I\bar{c}_{i\uparrow}(\tau)\bar{c}_{i\downarrow}(\tau)$
and obtain
\begin{eqnarray}
 &  & Z=Z_{0}^{B}\det\left[\mathbf{J}^{-1}\right]\int\mathcal{D}\left[\bar{c},c,\bar{\psi},\psi,\bar{\Delta},\Delta\right]\nonumber \\
 &  & \times e^{-\frac{1}{\hbar}\sum_{ij}\int_{0}^{\hbar\beta}d\tau J_{ij}^{-1}\left[\bar{\psi}_{i}(\tau)+I\bar{c}_{i\uparrow}(\tau)\bar{c}_{i\downarrow}(\tau)\right]\left[\psi_{j}(\tau)+Ic_{j\downarrow}(\tau)c_{j\uparrow}(\tau)\right]}\nonumber \\
 &  & \times e^{-\frac{1}{\hbar}\tilde{S}_{0}^{F}\left[\bar{c,}c,\bar{\Delta},\Delta\right]-\frac{1}{\hbar}W_{1}\left[\bar{\psi,}\psi\right]},\label{eq: effective_action_1}
\end{eqnarray}
where we define
\begin{equation}
W_{1}\left[\bar{\psi,}\psi\right]=-\hbar\ln\left\langle e^{-\frac{1}{\hbar}\sum_{i}\int_{0}^{\hbar\beta}d\tau\left(-\bar{\psi}_{i}(\tau)b_{i}(\tau)+c.c.\right)}\right\rangle _{0}^{B}.
\end{equation}
Within the strong-coupling approach ($J\ll U$) it is convenient to
expand $W_{1}\left[\bar{\psi,}\psi\right]$ in terms of $\psi_{i}\left(\tau\right)$,
$\bar{\psi}_{i}\left(\tau\right)$ fields, namely 
\begin{eqnarray}
 &  & W_{1}\left[\bar{\psi,}\psi\right]=\sum_{p=1}^{\infty}\frac{\left(-1\right)^{p}}{\left(p!\right)^{2}}\int_{0}^{\hbar\beta}d\tau_{1}...d\tau_{p}d\tau_{1}'...d\tau_{p}'\nonumber \\
 &  & \times\sum_{i}G_{i}^{p,c}(\tau{}_{1}',\,...,\,\tau'_{p},\,\tau{}_{1},\,...,\,\tau{}_{p})\nonumber \\
 &  & \times\bar{\psi}_{i}(\tau_{1}')...\bar{\psi}_{i}(\tau_{p}')\psi_{i}(\tau_{1})...\psi_{i}(\tau_{p}),
\end{eqnarray}
where $G_{i}^{p,c}(\tau{}_{1}',\,...,\,\tau'_{p},\,\tau{}_{1},\,...,\,\tau{}_{p})$
are connected local Green functions 
\begin{eqnarray}
 &  & G_{i}^{p,c}(\tau{}_{1}',\,...,\,\tau'_{p},\,\tau{}_{1},\,...,\,\tau{}_{p})\nonumber \\
 &  & \left.=\frac{\left(-1\right)^{p}\delta^{\left(2p\right)}W_{1}\left[\bar{\psi,}\psi\right]}{\delta\bar{\psi}_{i}(\tau_{1}')...\delta\bar{\psi}_{i}(\tau_{p}')\delta\psi_{i}(\tau_{1})...\delta\psi_{i}(\tau_{p})}\right|_{\bar{\psi}=\psi=0}.\label{eq: Gpc}
\end{eqnarray}
Then, truncating $W_{1}\left[\bar{\psi,}\psi\right]$ to quartic order
and inserting the results to Eq. (\ref{eq: effective_action_1}),
one gets the following effective action
\begin{eqnarray}
 &  & S^{eff}\left[\bar{c},c,\bar{\psi},\psi,\bar{\Delta},\Delta\right]\nonumber \\
 &  & =\tilde{S}_{0}^{B}\left[\bar{\psi,}\psi\right]+\sum_{ij}\int_{0}^{\hbar\beta}d\tau\left[\bar{\psi}_{i}(\tau)+I\bar{c}_{i\uparrow}(\tau)\bar{c}_{i\downarrow}(\tau)\right]\nonumber \\
 &  & \times J_{ij}^{-1}\left[\psi_{j}(\tau)+Ic_{j\downarrow}(\tau)c_{j\uparrow}(\tau)\right]+\tilde{S}_{0}^{F}\left[\bar{c,}c,\bar{\Delta},\Delta\right]\,\nonumber \\
 &  & -\frac{1}{4}\sum_{i}\int_{0}^{\hbar\beta}d\tau d\tau'd\tau''d\tau'''G_{i}^{2,c}\left(\tau,\tau',\tau'',\tau'''\right)\nonumber \\
 &  & \times\bar{\psi}_{i}(\tau''')\bar{\psi}_{i}(\tau'')\psi_{i}(\tau')\psi_{i}(\tau)\,.,\label{eq: pair hopping}
\end{eqnarray}
with
\begin{eqnarray}
\tilde{S}_{0}^{B}\left[\bar{\psi,}\psi\right] & = & \sum_{i}\int_{0}^{\hbar\beta}d\tau d\tau'G_{i}^{1,c}\left(\tau,\tau'\right)\bar{\psi}_{i}(\tau')\psi_{i}(\tau)
\end{eqnarray}
It is interesting to point out here that the pair hopping term naturally
emerges in the effective action from Eq. (\ref{eq: pair hopping}),
i.e. the term $I^{2}\sum_{ij}\int_{0}^{\hbar\beta}d\tau J_{ij}^{-1}\bar{c}_{i\uparrow}(\tau)\bar{c}_{i\downarrow}(\tau)c_{j\downarrow}(\tau)c_{j\uparrow}(\tau)$
and is induced by the resonant interaction $I$. 

Further, we perform the second HS transformation in terms of $J_{ij}^{-1}$
, i.e.
\begin{eqnarray}
 &  & -\sum_{ij}\int_{0}^{\hbar\beta}d\tau\left[\bar{\psi}_{i}(\tau)+I\bar{c}_{i\uparrow}(\tau)\bar{c}_{i\downarrow}(\tau)\right]\nonumber \\
 &  & \times J_{ij}^{-1}\left[\psi_{j}(\tau)+Ic_{j\downarrow}(\tau)c_{j\uparrow}(\tau)\right]\nonumber \\
 &  & \rightarrow\sum_{ij}\int_{0}^{\hbar\beta}d\tau J_{ij}\bar{\phi}_{i}(\tau)\phi_{j}(\tau)\nonumber \\
 &  & -\left\{ \sum_{i}\int_{0}^{\hbar\beta}d\tau\bar{\phi}_{i}(\tau)\left[\psi_{i}(\tau)+Ic_{i\downarrow}(\tau)c_{i\uparrow}(\tau)\right]+c.c.\right\} ,\label{eq: 2 HS}
\end{eqnarray}
where the new HS fields are $\phi_{i}(\tau)$, $\bar{\phi}_{i}(\tau)$. In comparison to the fields from the first HS (Eq. (\ref{eq: 1 HS})), the $\phi_{i}(\tau)$, $\bar{\phi}_{i}(\tau)$ fields have the same generating functional as the original $b_{i}(\tau)$, $\bar{b}_{i}(\tau)$ fields. Therefore using the $\phi_{i}(\tau)$, $\bar{\phi}_{i}(\tau)$ fields is more suitable in the physical analysis because their correlation functions have  the same interpretation as the correlation functions for the original $b_{i}(\tau)$, $\bar{b}_{i}(\tau)$ fields. To clarify this, in Appendix \ref{sub: generating functional}, we add the proof that both fields have the same generating functional. Moreover, beyond this useful fact about $\phi_{i}(\tau)$, $\bar{\phi}_{i}(\tau)$, it is worth mentioning here that these fields, in the limit of BHM (when $I=0$), yield properly normalized density of states in the BHM superfluid phase \cite{2005PhRvA..71c3629S} (properties of the SF spectrum in the full BFHM need further studies).

After applying second HS (Eq. (\ref{eq: 2 HS})) to the Eq. (\ref{eq: pair hopping}), corresponding effective action is 
\begin{eqnarray}
 &  & S^{eff}\left[\bar{c},c,\bar{\phi,}\phi,\bar{\Delta},\Delta\right]\nonumber \\
 &  & =-\sum_{ij}\int_{0}^{\hbar\beta}d\tau J_{ij}\bar{\phi}_{i}(\tau)\phi_{j}(\tau)\nonumber \\
 &  & +\left\{ I\sum_{i}\int_{0}^{\hbar\beta}d\tau\bar{\phi}_{i}(\tau)c_{i\downarrow}(\tau)c_{i\uparrow}(\tau)+c.c.\right\} \nonumber \\
 &  & +\tilde{S}_{0}^{F}\left[\bar{c,}c,\bar{\Delta},\Delta\right]+W_{2}\left[\bar{\phi,}\phi\right]\,,
\end{eqnarray}
with denotation\begin{widetext}
\begin{equation}
W_{2}\left[\bar{\phi,}\phi\right]=-\hbar\ln\left\langle e^{-\frac{1}{\hbar}\sum_{i}\int_{0}^{\hbar\beta}d\tau\left(\bar{\phi}_{i}(\tau)\psi_{i}(\tau)+c.c.\right)+\frac{1}{4}\sum_{i}\int_{0}^{\hbar\beta}d\tau d\tau'd\tau''d\tau'''G_{i}^{2,c}\left(\tau,\tau',\tau'',\tau'''\right)\bar{\psi}_{i}(\tau''')\bar{\psi}_{i}(\tau'')\psi_{i}(\tau')\psi_{i}(\tau)}\right\rangle _{0}^{B,eff},
\end{equation}
and where the statistical average $\left\langle ...\right\rangle _{0}^{B,eff}$
is defined as $\left(\tilde{Z}_{0}^{B}\right)^{-1}\int\mathcal{D}\left[\bar{\psi},\psi\right]...e^{-\tilde{S}_{0}^{B}/\hbar}$
with $\tilde{Z}_{0}^{B}=\int\mathcal{D}\left[\bar{\psi},\psi\right]e^{-\tilde{S}_{0}^{B}/\hbar}$.
And once again truncating $W_{2}\left[\bar{\phi,}\phi\right]$ to
the quartic order and retaining only the terms which are not ``anomalous''
\cite{Dupuis2001,2005PhRvA..71c3629S,PhysRevA.84.033620,FitzpatrickKennet}, we obtained
the final form of statistical sum $\tilde{Z}^{eff}$ with effective
action $\tilde{S}^{eff}$ (in which the fermionic degrees of freedom
were integrated out), i.e.
\begin{equation}
\tilde{Z}^{eff}=\int\mathcal{D}\left[\bar{\phi},\phi,\bar{\Delta},\Delta\right]e^{-\frac{1}{\hbar}\tilde{S}^{eff}\left[\bar{\phi},\phi,\bar{\Delta},\Delta\right]},
\end{equation}
\begin{eqnarray}
 &  & \tilde{S}^{eff}\left[\bar{\phi},\phi,\bar{\Delta},\Delta\right]=-Tr\ln\left(-G_{F}^{-1}\left(i,\, j,\,\tau\right)\right)+\frac{1}{V}\sum_{i}\left|\Delta_{i}(\tau)\right|^{2}-\sum_{ij}\int_{0}^{\hbar\beta}d\tau J_{ij}\bar{\phi}_{i}(\tau)\phi_{j}(\tau)\nonumber \\
 &  & -\sum_{i}\int_{0}^{\hbar\beta}d\tau d\tau'\left[G_{i}^{1,c}\left(\tau,\tau'\right)\right]^{-1}\bar{\phi}_{i}(\tau')\phi_{i}(\tau)+\frac{1}{4}\sum_{i}\int_{0}^{\hbar\beta}d\tau d\tau'd\tau''d\tau'''\Gamma_{i}^{2,c}\left(\tau,\tau',\tau'',\tau'''\right)\bar{\phi}_{i}(\tau''')\bar{\phi}_{i}(\tau'')\phi_{i}(\tau')\phi_{i}(\tau),\label{eq: effective action 2}
\end{eqnarray}

\end{widetext}where we introduced the matrix fermionic Green function
\begin{eqnarray}
 &  & G_{F}^{-1}\left(i,\, j,\,\tau\right)=\nonumber \\
 &  & =\left[\begin{array}{cc}
\left(-\hbar\frac{\partial}{\partial\tau}+\mu\right)\delta_{ij}+t_{ij} & \Delta_{i}(\tau)-I\phi_{i}(\tau)\\
\bar{\Delta}_{i}(\tau)-I\bar{\phi}_{i}(\tau) & \left(-\hbar\frac{\partial}{\partial\tau}-\mu\right)\delta_{ij}-t_{ij}
\end{array}\right],
\end{eqnarray}
and effective interaction between bosons
\begin{eqnarray}
 &  & \Gamma_{i}^{2,c}\left(\tau,\tau',\tau'',\tau'''\right)\nonumber \\
 &  & \left.=\frac{\delta^{\left(4\right)}W_{2}\left[\bar{\phi,}\phi\right]}{\delta\bar{\phi}_{i}(\tau_{1}')\delta\bar{\phi}_{i}(\tau_{2}')\delta\phi_{i}(\tau_{1})\delta\phi_{i}(\tau_{2})}\right|_{\bar{\phi}=\phi=0}.\label{eq: Gamma_pc}
\end{eqnarray}
In the following, to analyze the phase diagrams of BFHM, we focus
on the saddle point approximation for the effective action from Eq.
(\ref{eq: effective action 2}). Moreover, we point out that this
effective action could be also used as a starting point for more general
considerations which include the fluctuations around saddle point
approximation. Formally, it can be performed by expanding $G_{F}^{-1}\left(i,\, j,\,\tau\right)$
in terms of $\Delta_{i}(\tau)-I\phi_{i}(\tau)$ fields.

\subsection{Saddle point approximation of the effective action \label{sub:Saddle-point-of the effective action}}

To investigate the phase diagram which is described by the BFHM effective
action from Eq. (\ref{eq: effective action 2}), we apply the mean-field
type approximations.

At first, we rewrite Eq. (\ref{eq: effective action 2}) in the Matsubara
frequencies ($\omega_{m}$, $\nu_{n}$) and wave vector ($\mathbf{k}$,
$\mathbf{q}$, $\mathbf{p}$) representation, which results in $c_{i}(\tau)\rightarrow c_{\mathbf{k}m}$,
$\phi_{i}(\tau)\rightarrow\phi_{\mathbf{q}n}$, $\Delta_{i}(\tau)\rightarrow\Delta_{\mathbf{q}n}$.
The Matsubara frequencies are defined as $\omega_{m}=\left(2m+1\right)\pi/\beta$
and $\nu_{n}=2n\pi/\beta$ where $m,\, n\in\mathbb{Z}$. Then, applying
the Bogoliubov like substitution to the $\phi_{\mathbf{0}0}$ and
$\Delta_{\mathbf{0}0}$ components, i.e. $\phi_{\mathbf{0}0}\rightarrow\sqrt{N\hbar\beta}\phi_{0}$
and $\Delta_{\mathbf{0}0}\rightarrow\sqrt{N\hbar\beta}\Delta_{0}$
and omitting the fluctuating bosonic parts $\Delta_{\mathbf{q}n}$
and $\phi_{\mathbf{q}n}$, the mean-field effective action is obtained,
i.e.
\begin{eqnarray}
 &  & S_{MF}^{eff}=\left\{ \epsilon_{\mathbf{0}}-\hbar\left[G^{1,c}\left(i\nu_{n}=0\right)\right]^{-1}\right\} N\hbar\beta\left|\phi_{0}\right|^{2}\nonumber \\
 &  & +\frac{g}{2}\left(N\hbar\beta\right)^{2}\left|\phi_{0}\right|^{4}+\frac{N\hbar\beta}{V}\left|\Delta_{0}\right|^{2}\nonumber \\
 &  & -\mathrm{Tr}\ln\left(-N\beta G_{F}^{-1}\left(i\omega_{m},\,\mathbf{k}\right)\right)\,,\label{eq: effective action 3}
\end{eqnarray}
where
\begin{equation}
G_{F}^{-1}\left(\mathbf{k},\, i\nu_{m}\right)=\left[\begin{array}{cc}
i\hbar\omega_{m}-\xi_{\mathbf{k}} & \Delta_{0}-I\phi_{0}\\
\bar{\Delta}_{0}-I\bar{\phi}_{0} & i\hbar\omega_{m}+\xi_{\mathbf{k}}
\end{array}\right]\,,
\end{equation}
with $\epsilon_{\mathbf{q}}=-2J\sum_{\alpha}\cos q_{\alpha}$, $\xi_{\mathbf{k}}=t_{\mathbf{k}}-\mu$,
$t_{\mathbf{k}}=-2t\sum_{\alpha}\cos k_{\alpha}$ (symbol $\alpha\in\left\{ x,\, y,\, z\right\} $ denotes Cartesian coordinates). Moreover, in further calculations we also define coordinate number $z=6$ which is related to the $\epsilon_{\mathbf{q}}$ by expression $\epsilon_{\mathbf{0}}=-Jz$. Here, we restrict our consideration to the simple cubic lattices. 
The explicit form of $G^{1,c}\left(i\nu_{n}\right)$ is given in Appendix
\ref{sub:Local-Green's-function}. Moreover, in Eq. (\ref{eq: effective action 3})
we use static approximation to the $\Gamma_{i}^{2,c}$ function and
denote this limit by $2g$ (here we do not use the explicit form of
$g$ but it could be found in Ref. \cite{2005PhRvA..71c3629S}). 

To describe the ordered phase in terms of $\phi_{0}$ and $\Delta_{0}$
we calculate the saddle point of the above effective action 
\begin{equation}
\frac{\partial}{\partial\bar{b}_{0}}S_{MF}^{eff}=0\,,
\end{equation}
\begin{equation}
\frac{\partial}{\partial\bar{\Delta}_{0}}S_{MF}^{eff}=0\,.
\end{equation}
This results in the following coupled equations\begin{widetext}
\begin{equation}
\left\{ \begin{array}{l}
\left\{ \epsilon_{\mathbf{0}}-\hbar\left[G^{1,c}\left(i\nu_{n}=0\right)\right]^{-1}\right\} \phi_{0}+gN\hbar\beta\left|\phi_{0}\right|^{2}\phi_{0}=-\frac{I}{N\hbar\beta}\sum_{m\mathbf{k}}G_{F}^{12}\left(\mathbf{k},\, i\hbar\omega_{m}\right)=-\frac{I}{N}\sum_{\mathbf{k}}\frac{\left(Vx_{0}-I\phi_{0}\right)}{2E_{\mathbf{k}}}\tanh\left(\frac{\beta}{2}E_{\mathbf{k}}\right),\\
x_{0}=\frac{1}{N\hbar\beta}\sum_{m\mathbf{k}}G_{F}^{12}\left(\mathbf{k},\, i\hbar\omega_{m}\right)=\frac{1}{N}\sum_{\mathbf{k}}\frac{\left(Vx_{0}-I\phi_{0}\right)}{2E_{\mathbf{k}}}\tanh\left(\frac{\beta}{2}E_{\mathbf{k}}\right),
\end{array}\right.\label{eq: equation coupled}
\end{equation}
\end{widetext}where  $Vx_{0}=\Delta_{0}$ and
\begin{equation}
E_{\mathbf{k}}=\sqrt{\xi_{\mathbf{k}}^{2}+\left|I\phi_{0}-Vx_{0}\right|^{2}}\,.\label{eq: EkF}
\end{equation}
From Eqs. (\ref{eq: equation coupled}) one immediately sees that
$x_{0}$ and $\phi_{0}$ are non-linearly coupled to each other,
i.e. 
\begin{equation}
\left\{ \epsilon_{\mathbf{0}}-\hbar\left[G^{1,c}\left(i\nu_{n}=0\right)\right]^{-1}\right\} \phi_{0}+gN\hbar\beta\left|\phi_{0}\right|^{2}\phi_{0}=-Ix_{0}.
\end{equation}
which suggests that there is only one phase transition from the superfluid phase to normal phase.

Moreover, it is interesting to point out here, that above equation
correctly recovers the limiting cases of non-interacting ($U=0$)
and hard core ($U\rightarrow\infty$) bosons (in which fermionic interaction can be finite i.e. $V\neq0$). For $U=0$ the term
with $g$ disappears and one has $\hbar\left[G^{1,c}\left(i\nu_{n}=0\right)\right]^{-1}=\mu^{*}$,
therefore 
\begin{equation}
\phi_{0}=\frac{-I}{\epsilon_{\mathbf{0}}-\left(2\mu-2\Delta_{B}\right)}x_{0}\,,
\end{equation}
which corresponds to the well-known result without a lattice \cite{PhysRevA.67.063612}.
For $U\rightarrow\infty$, two Fock states are taken in Eq. (\ref{eq: G0}),
i.e. $n_{0}=0,\,1$, which gives $\hbar\left[G^{1,c}\left(i\nu_{n}=0\right)\right]^{-1}=\mu^{*}/\left(1-2n_{B,0}\right)$
with $n_{B,0}=e^{\beta\mu^{*}}/(1+e^{\beta\mu^{*}})$. Therefore,
for the hard-core bosons case one gets
\begin{equation}
\phi_{0}=\left(Ix_{0}+\epsilon_{\mathbf{0}}\phi_{0}\right)\frac{1-2n_{B,0}}{2\mu-2\Delta_{B}}\,,\label{eq: order paramter - HC}
\end{equation}
where we neglect the contribution from $g$ term by assuming a limit
of small order parameter $\phi_{0}$. This result (Eq. (\ref{eq: order paramter - HC}))
recovers the previous one from Ref. \cite{PhysRevB.36.180}.

We have also confirmed that Eqs. (\ref{eq: equation coupled}), in
the limit of small amplitude of $\phi_{0}$ (in which the term proportional
to $g$ could be neglected), can be recovered from the mean-field
and linear response considerations, see Appendix \ref{sub: meanfield - lienar response}.
Therefore, these both approaches lead to the same equation for critical
line considered in the rest of the paper.

At the end of this subsection, it is worth pointing out that the results, obtained in Secs. \ref{sec: effective action} and \ref{sub:Saddle-point-of the effective action}, are quite general and can be used for further analytical and numerical considerations in which $I$, $U$ and $V$ interactions are finite quantities. These results are interested on its own right and can be applied to study of e.g. superfluidity or critical phenomena. 
In our further analysis, we focus on the specific physical regime of derived theory in which BHM is set as our reference point.

\subsection{Phase diagram \label{sec: phase diagraaam}}
In this work, we are interested in the phase diagram of strongly correlated
bosonic regime ($J\ll U$). Therefore,
at the phase boundary where $x_{0}\rightarrow0,\:\phi_{0}\rightarrow0$
in Eqs. (\ref{eq: equation coupled}), critical line is obtained from
\begin{eqnarray}
\epsilon_{\mathbf{0}}-\hbar [G^{1,c}\left(i\nu_{n}=0\right)]^{-1}=\frac{I^{2}\Pi(T_c)}{1-V\Pi(T_c)}\label{eq:critical-line-equation-with-V},
\end{eqnarray}
where
\begin{eqnarray}
\Pi(T_c)=\frac{1}{N}\sum_{\mathbf{k}}\frac{1}{2\xi_{\mathbf{k}}}\tanh\left(\frac{\xi_{\mathbf{k}}}{2k_B T_c}\right).
\end{eqnarray}
It is interesting to notice here that in the case of $I=0$, the Eq. (\ref{eq:critical-line-equation-with-V}) and the equation in the second line of (\ref{eq: equation coupled}), get the forms which are known in the phase diagram analysis of BHM and BCS systems, respectively.

However, in our furhter analysis, we limit considerations to the case of $V=0$ for simplicity. Therefore we focus on the paring mechanism of fermions which comes from the converting interactions $I$. Then, by direct substitution of $V=0$ to  the Eq. (\ref{eq:critical-line-equation-with-V}), the phase boundary in BFHM is obtained from the equation
\begin{equation}
\epsilon_{\mathbf{0}}-\hbar [G^{1,c}\left(i\nu_{n}=0\right)]^{-1}=I^{2}\Pi(T_c).\label{eq:critical-line-equation}
\end{equation}

In further discussion we set $\hbar=1$ and $k_{B}=1$ for simplicity.

\begin{figure}[th]
\includegraphics[scale=0.5]{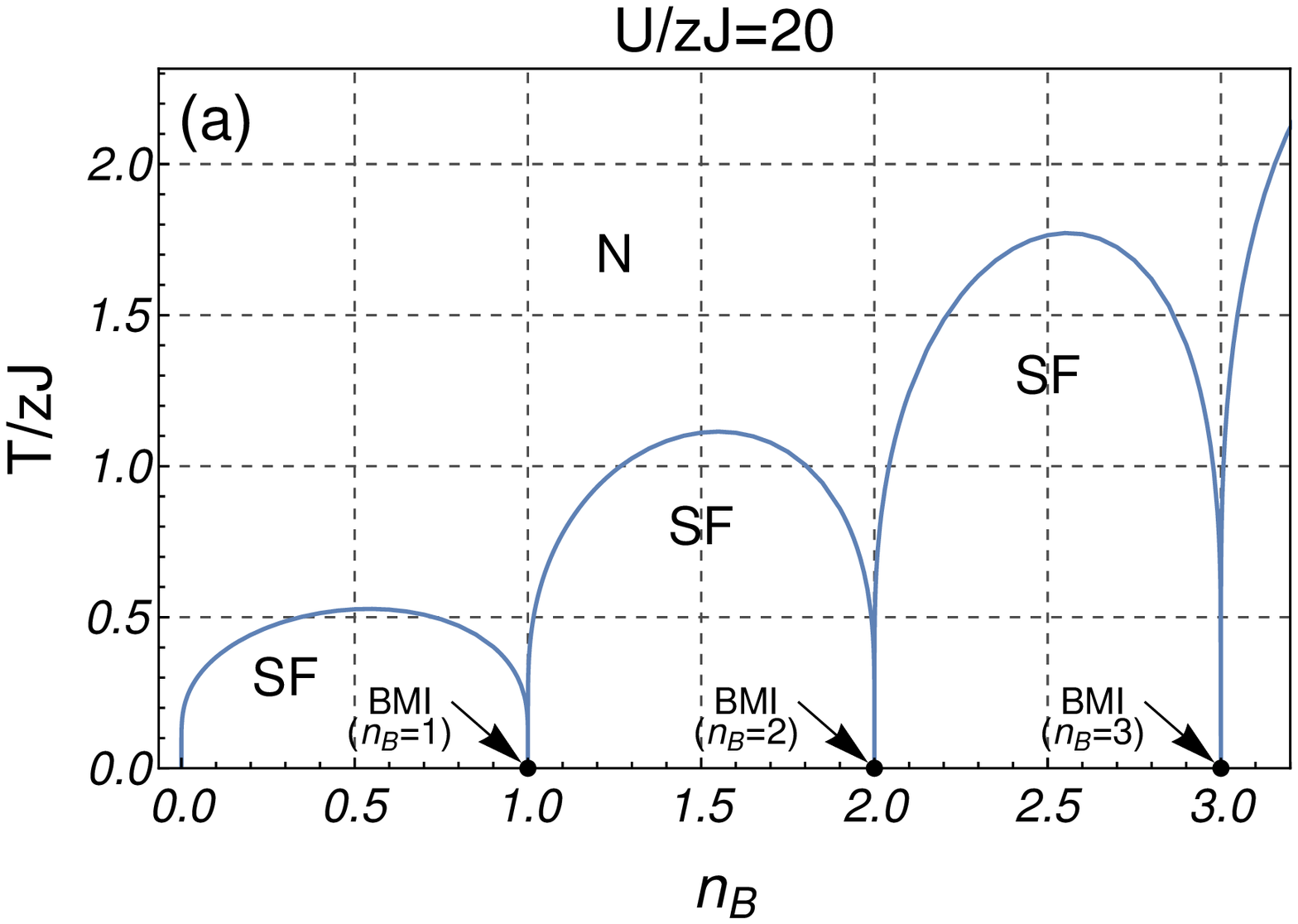}

\includegraphics[scale=0.5]{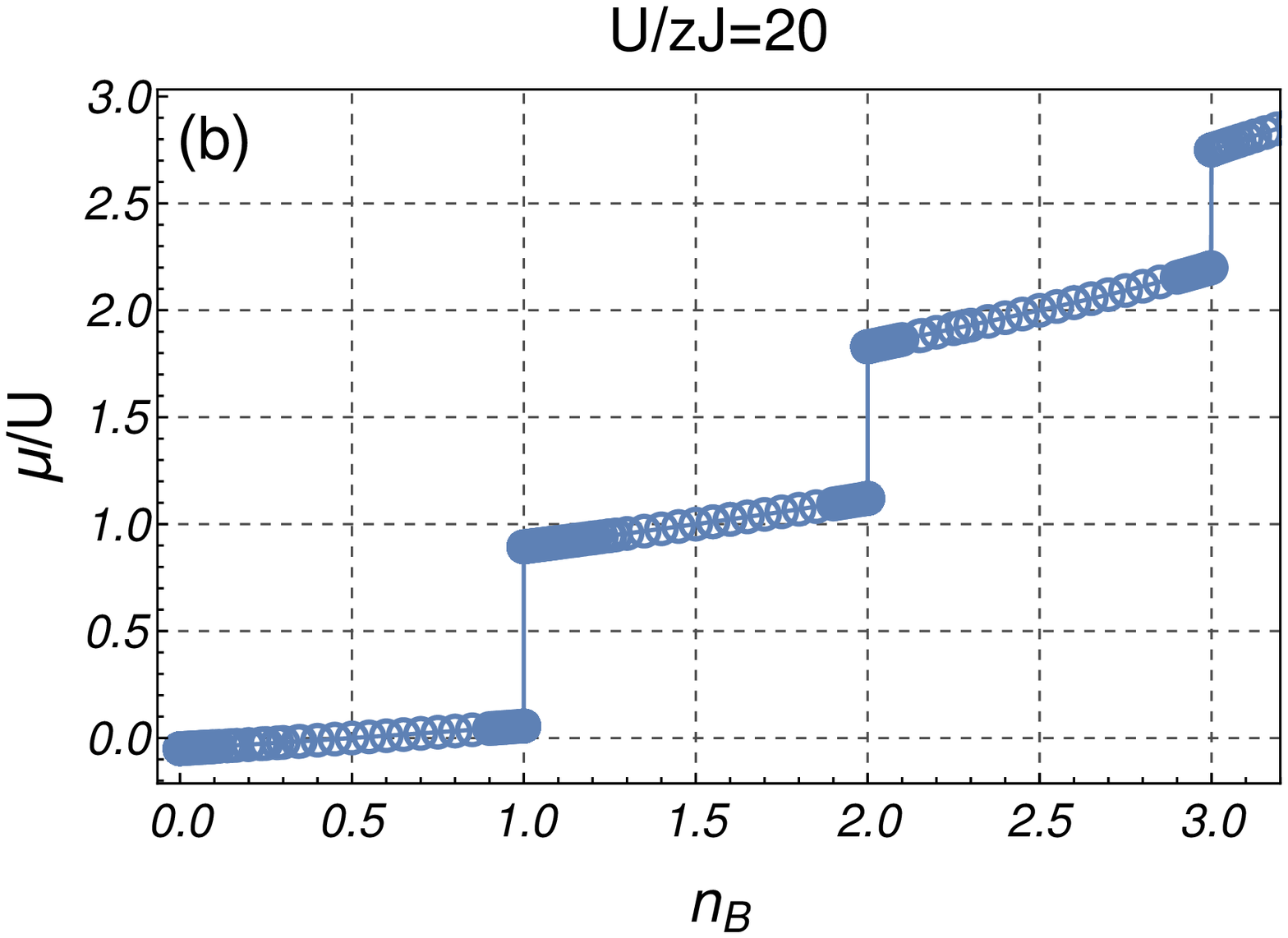}

\caption{(a) Mean-field phase diagram of BHM (temperature $T/Jz$ versus average
particle number per site $n_{B}$). (b) Chemical potential $\mu/U$
versus $n_{B}$ calculated along the critical line from Fig. (a).
For clarity, the circles are added on the numerical data points in
Fig. (b).\label{fig: pure BHM}}
\end{figure}

\begin{figure}[th]
\includegraphics[scale=0.65]{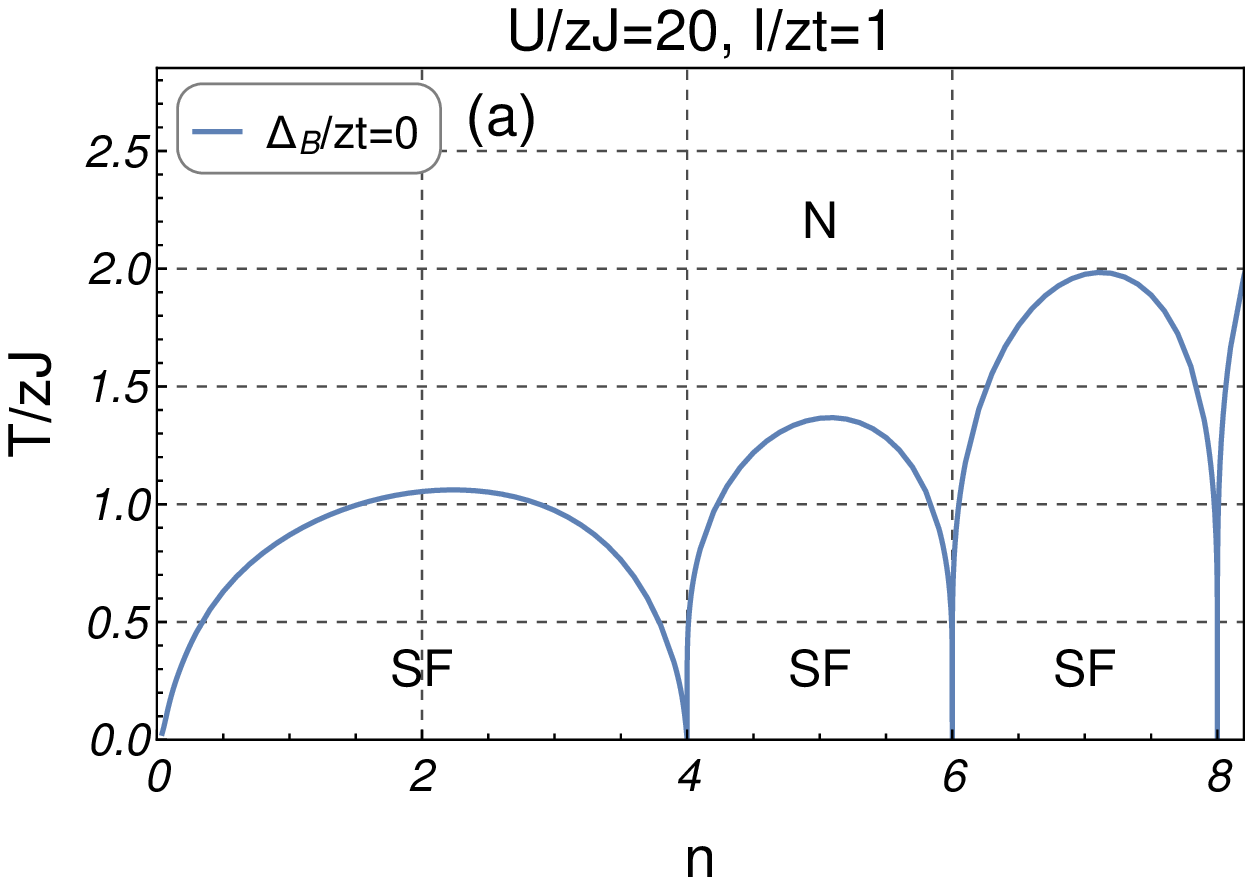}

\includegraphics[scale=0.32]{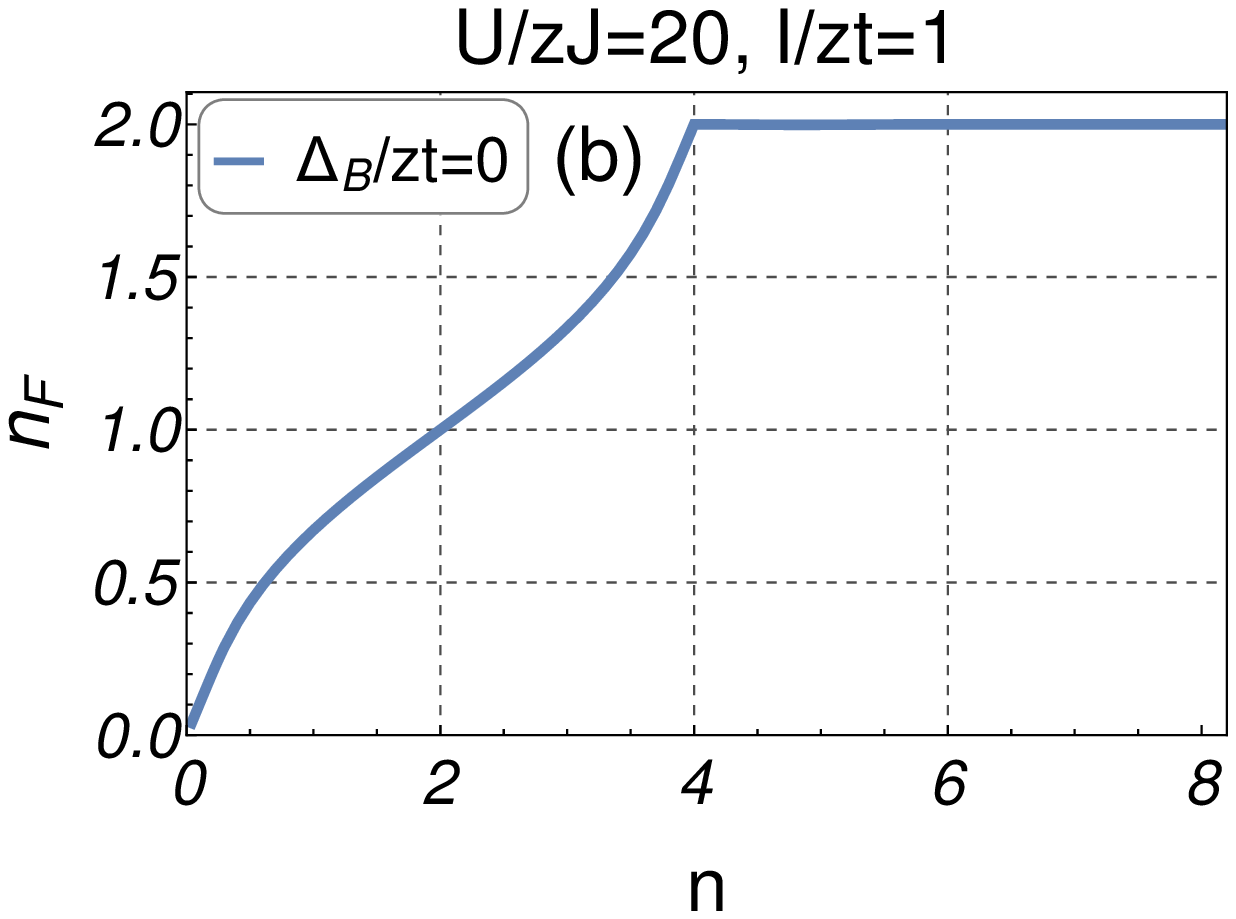}\includegraphics[scale=0.32]{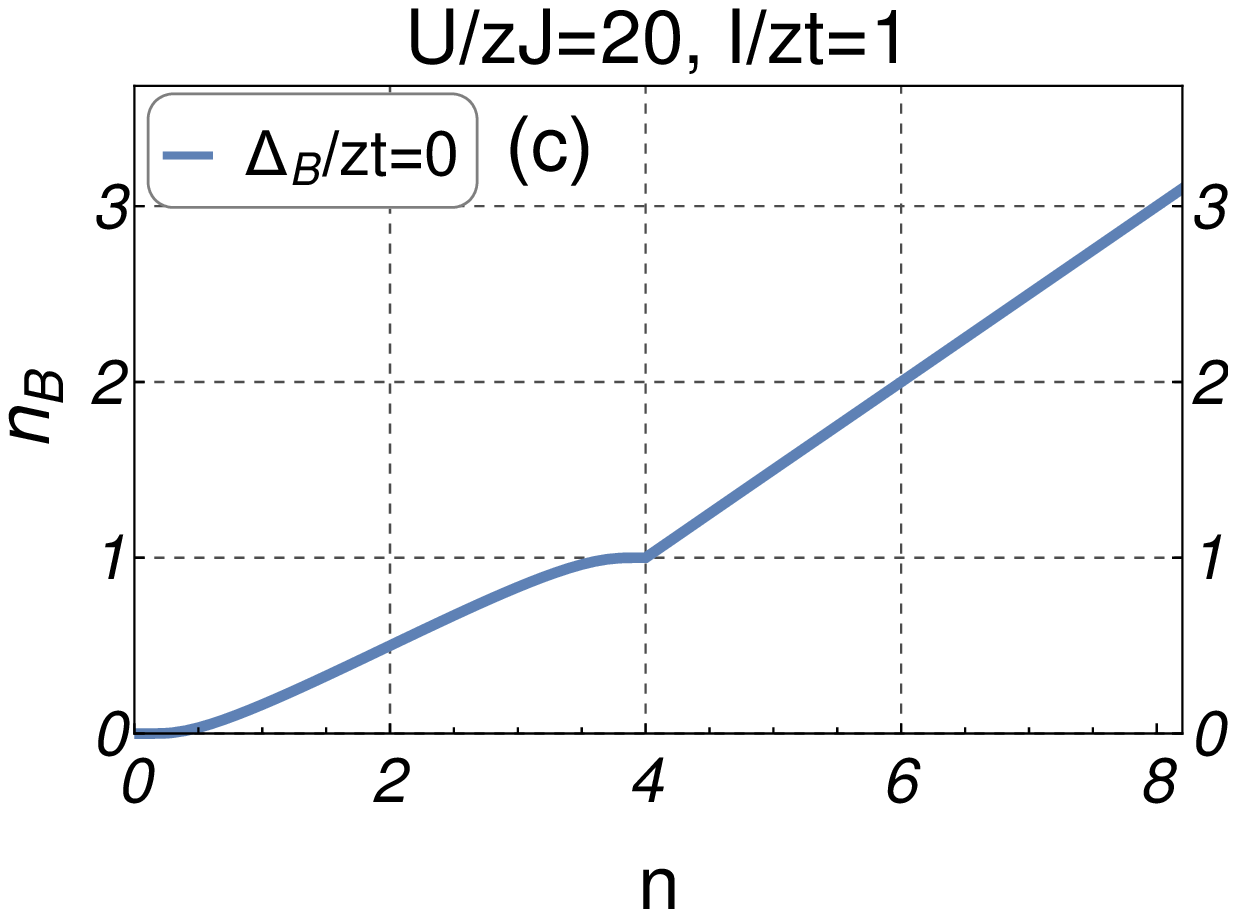}

\includegraphics[scale=0.32]{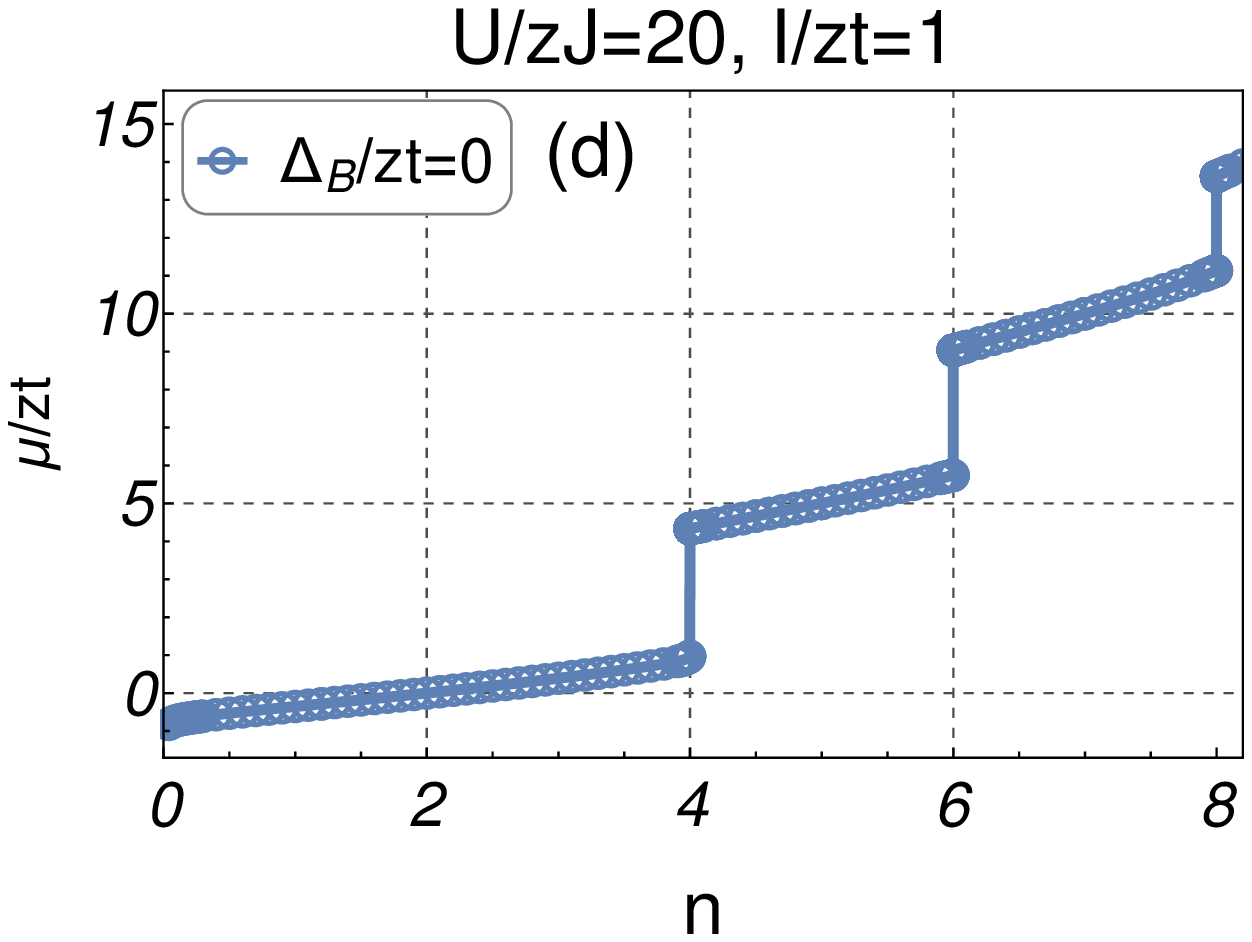}\includegraphics[scale=0.32]{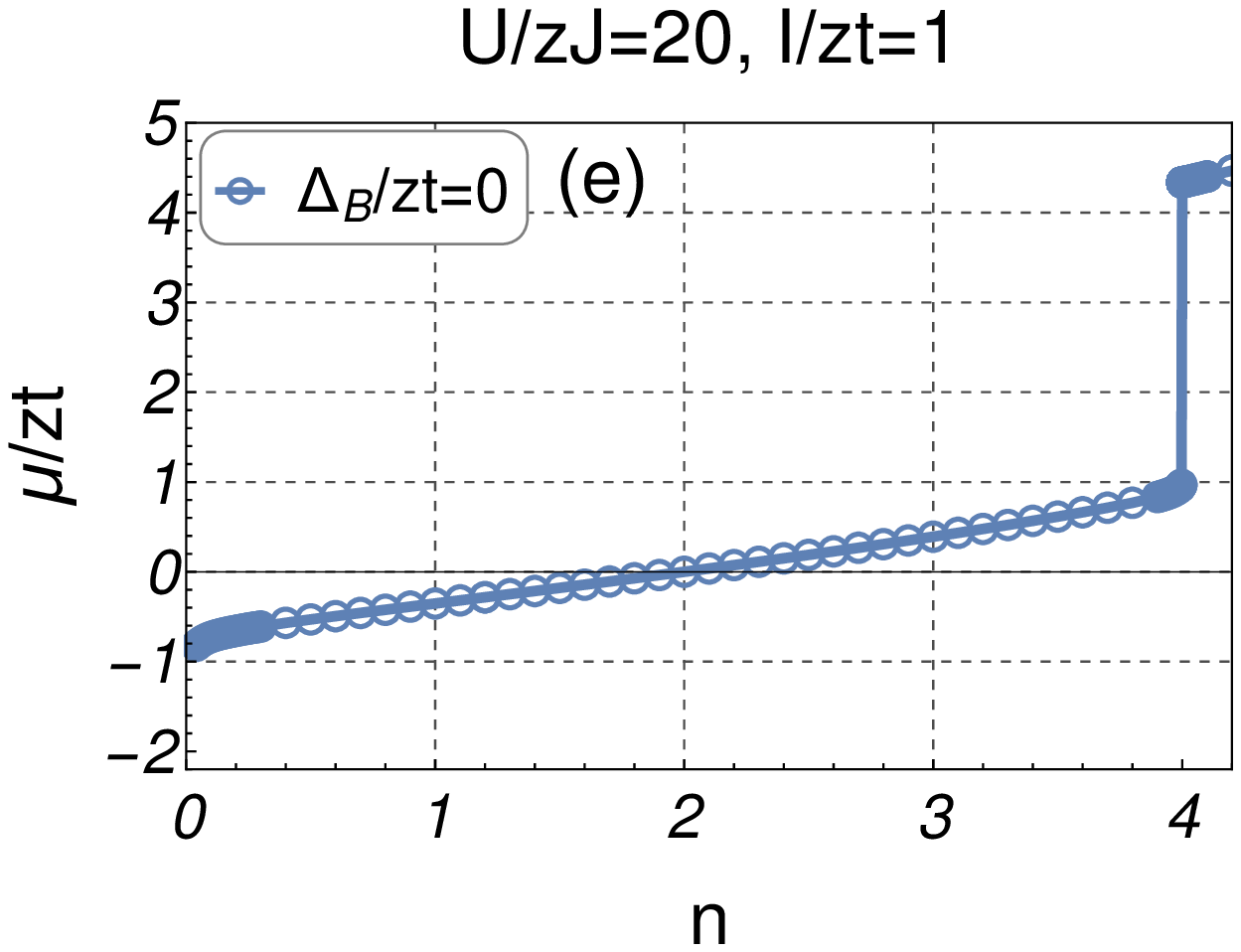}

\caption{(a) Finite temperature mean-field phase diagram of BFHM versus total
particle number $n=2n_{B}+n_{F}$ for zero detuning of parameter $\Delta_{B}$.
Figures (b), (c), (d) are plots of $n_{F}$, $n_{B}$, $\mu/zt$ versus
$n$, respectively (the data obtained are evaluated along the critical
line from Fig. (a)). Figure (e) is an enlargement of the vicinity
of zero chemical potential from plot (d). Plots are made assuming
that $U/zJ=20$, $I/zt=1$, $J=t/2$. For clarity, the circles are
added on the numerical data points in Figs. (d) and (e). \label{fig: Delta=00003D0}}
\end{figure}

\begin{figure}
\includegraphics[scale=0.65]{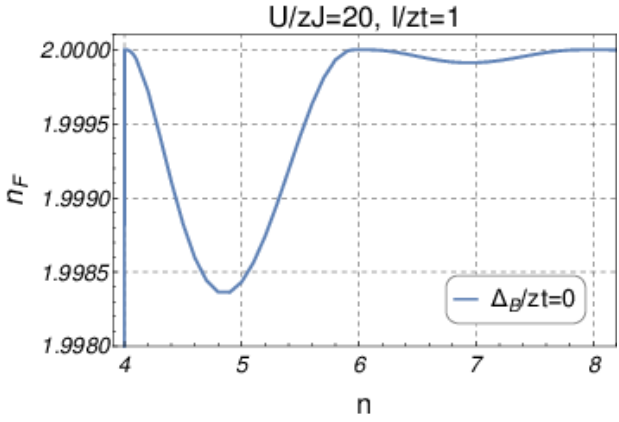}
\caption{Dependence of fermionic density $n_F$ on the total particle density $n$. This figure is an enlargement of the $n\in(4,8)$ region from Fig. \ref{fig: Delta=00003D0} b.  \label{fig: supplementtoFig2}}
\end{figure}

\begin{figure}[th]
\includegraphics[scale=0.65]{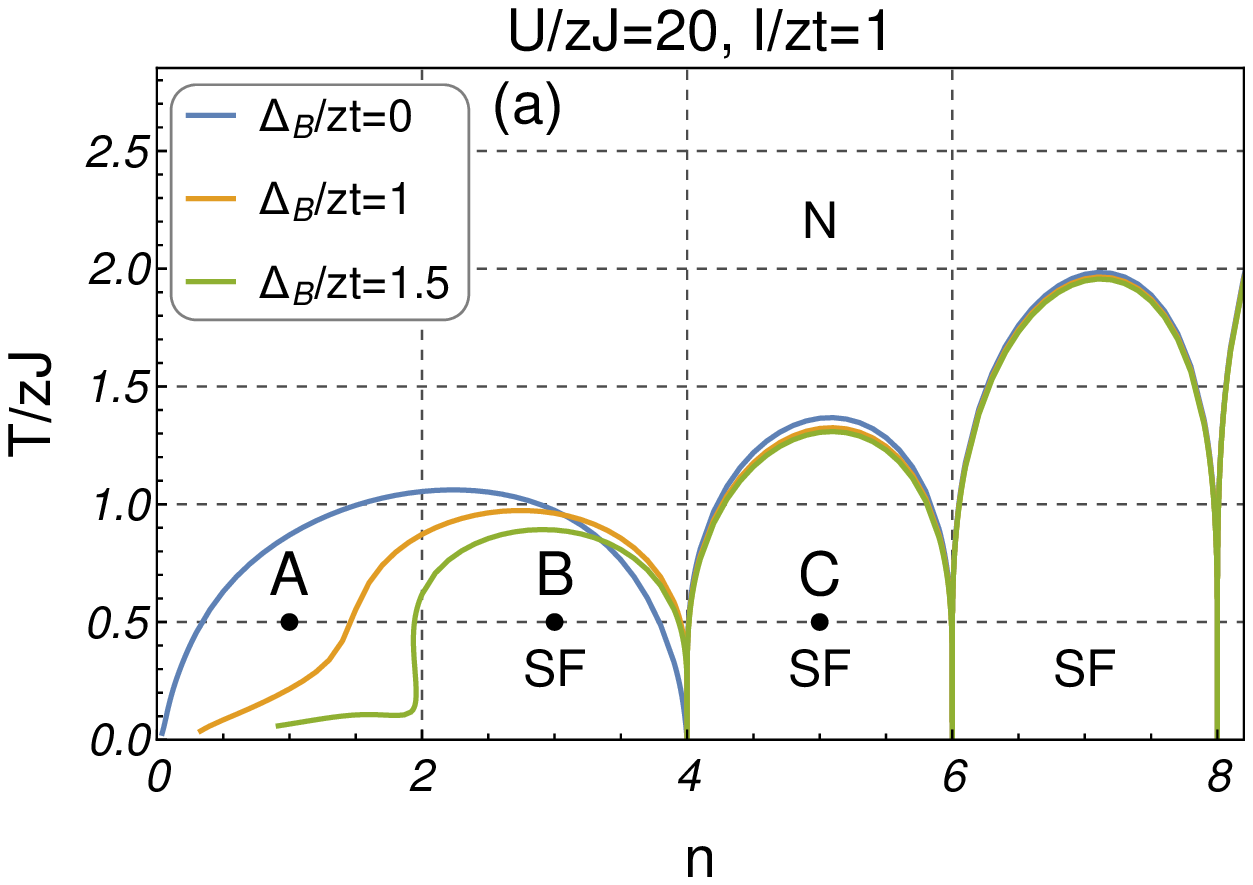}

\includegraphics[scale=0.32]{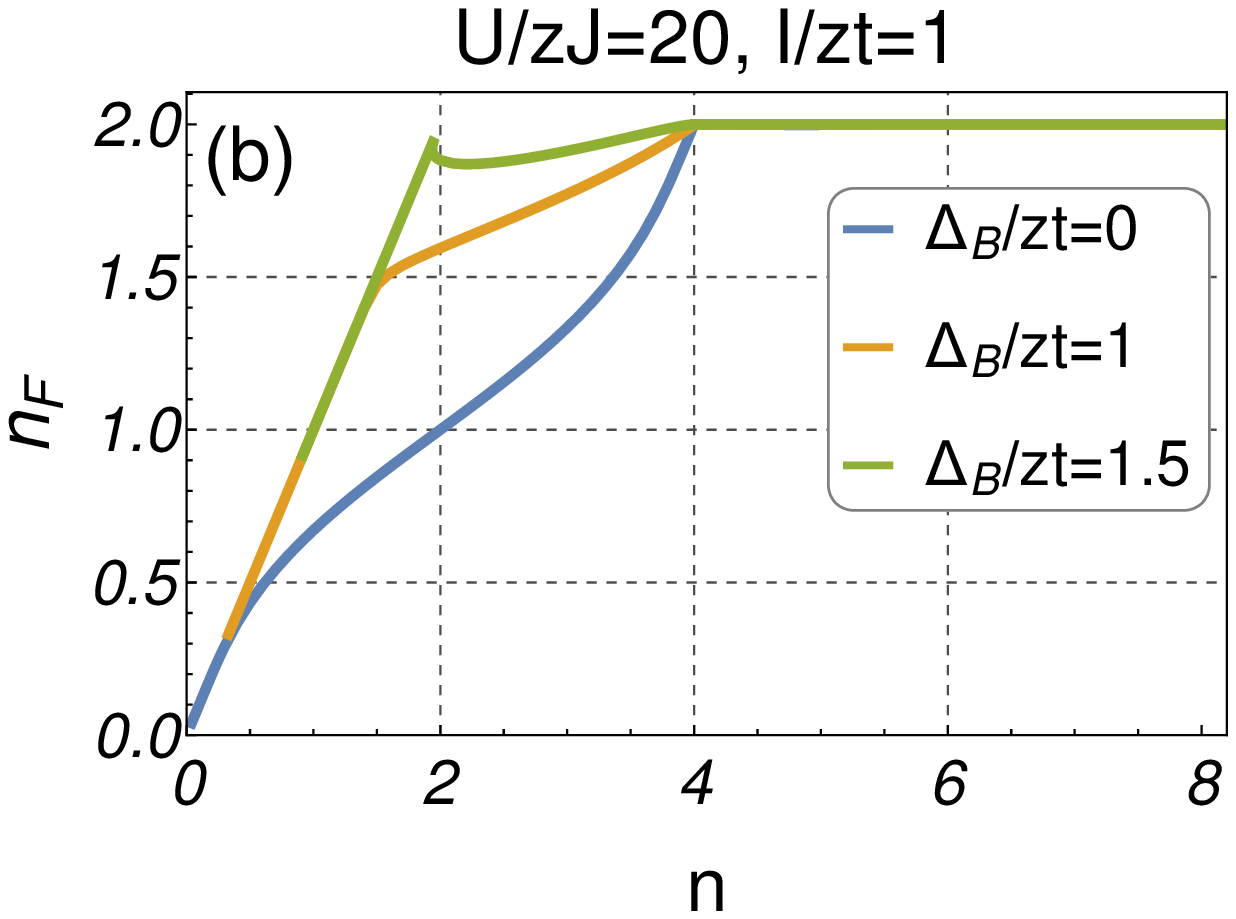}\includegraphics[scale=0.32]{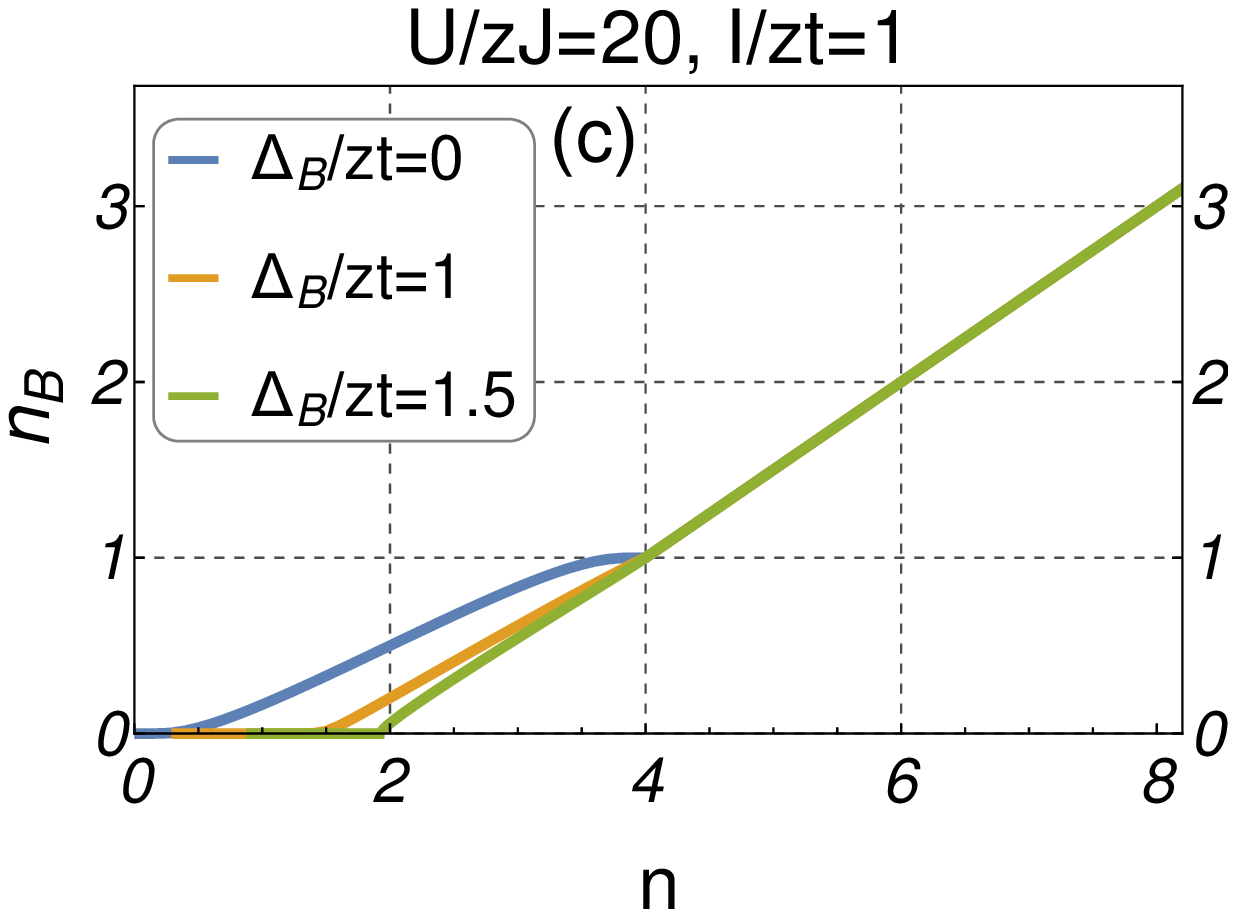}

\includegraphics[scale=0.32]{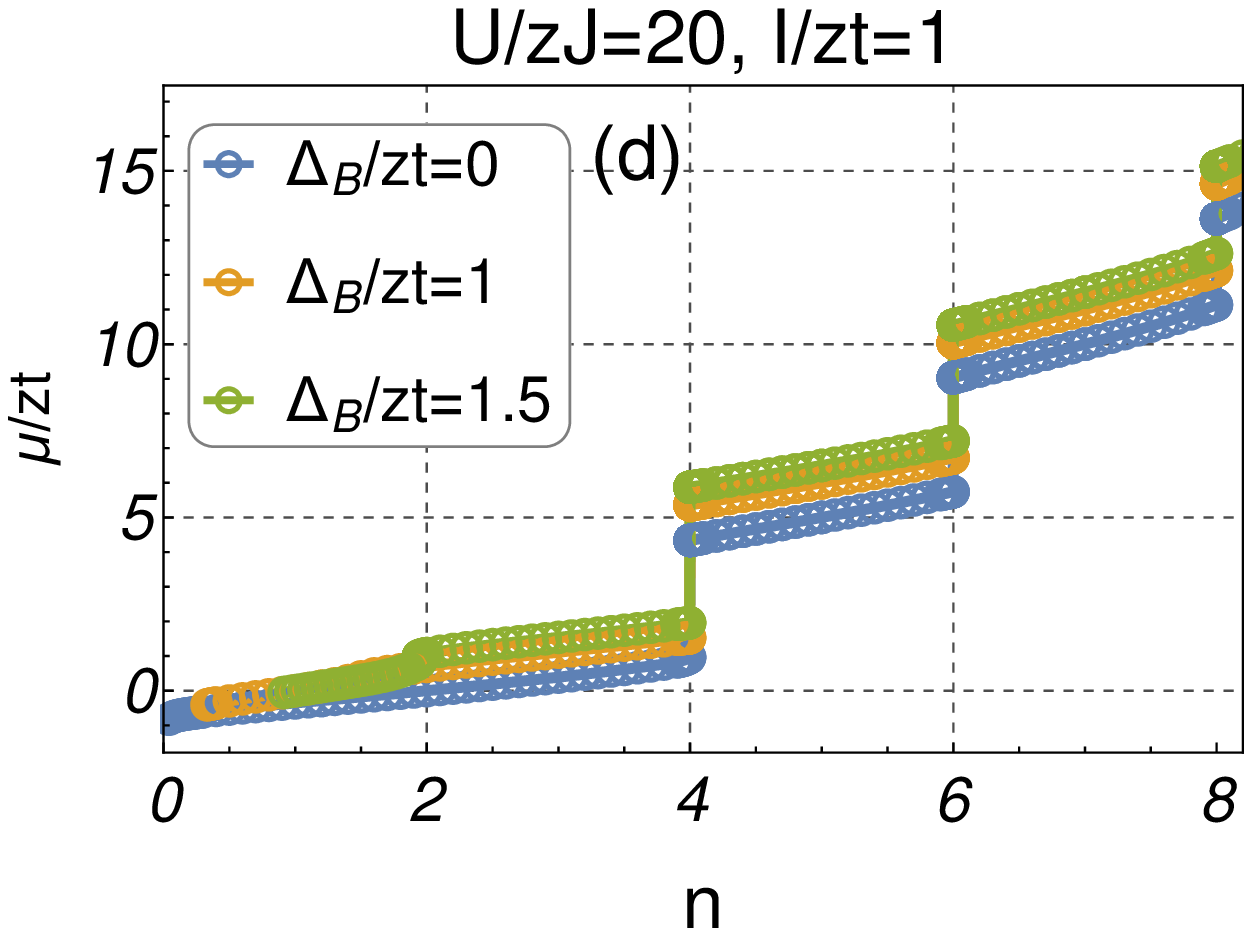}\includegraphics[scale=0.32]{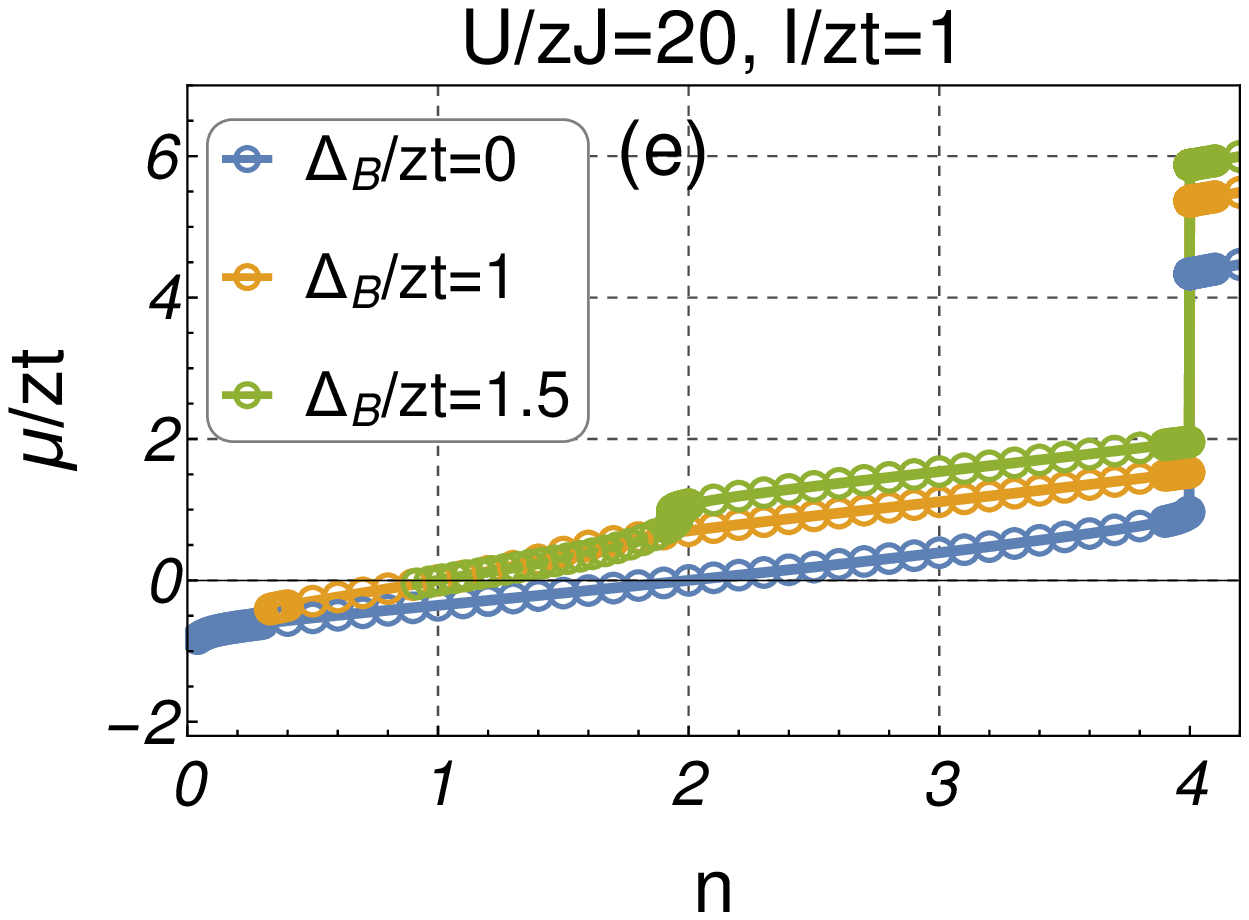}

\caption{(a) Finite temperature mean-field phase diagram of BFHM versus total
particle number $n=2n_{B}+n_{F}$ for different strengths of detuning
$\Delta_{B}$ (see legend). Figures (b), (c), (d) are plots of $n_{F}$,
$n_{B}$, $\mu/zt$ versus $n$, respectively (the data obtained are
evaluated along the critical line from Fig. (a)). Figure (e) is an
enlargement of the vicinity of zero chemical potential from plot (d).
Plots are made assuming that $U/zJ=20$, $I/zt=1$, $J=t/2$. For comparison,
we plot $\Delta_{B}/zt=0$ from Fig. \ref{fig: Delta=00003D0}. For
clarity, the circles are added on the numerical data points in Figs.
(d) and (e). Meaning of A, B and C points is given in Sec. \ref{sub:experiment}. \label{fig: positive detuning}}
\end{figure}

\begin{figure}[th]
\includegraphics[scale=0.65]{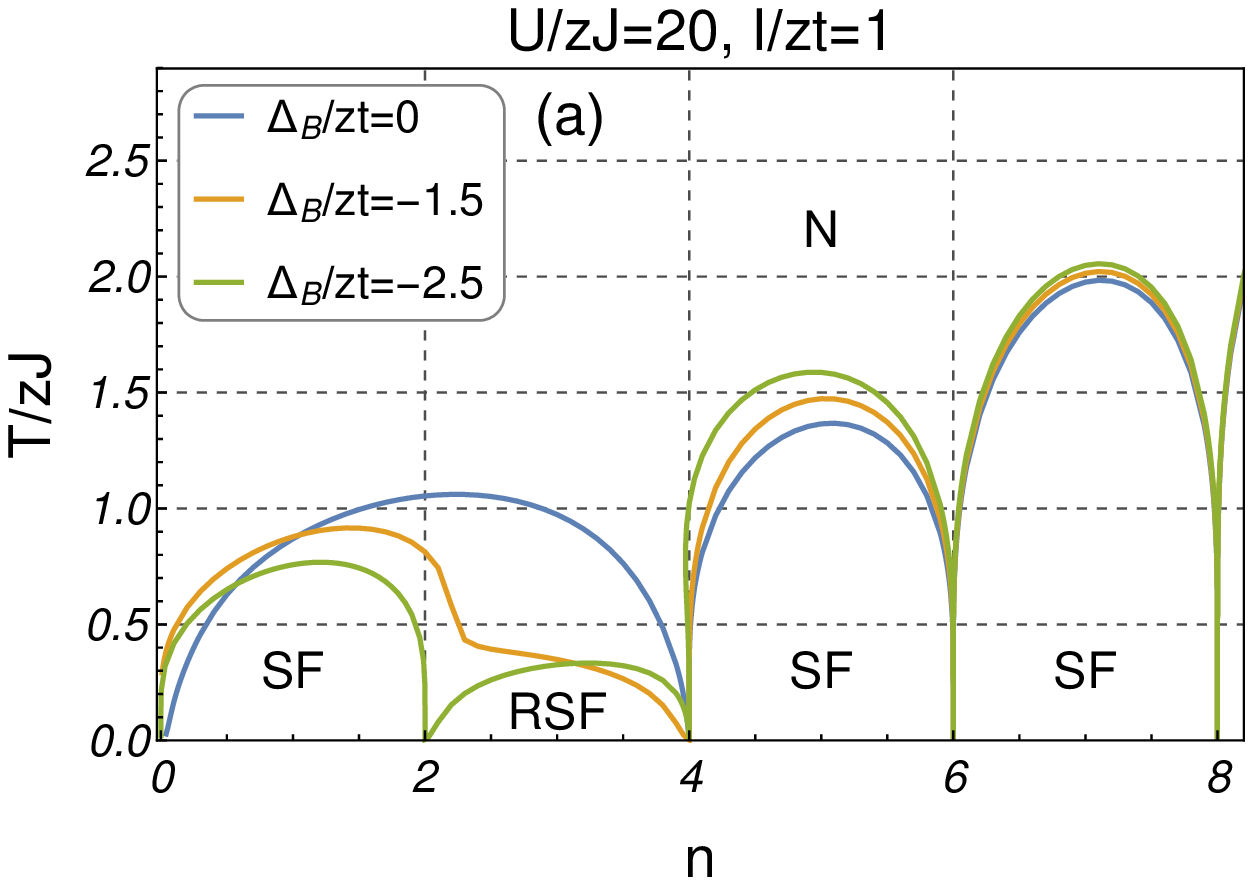}

\includegraphics[scale=0.32]{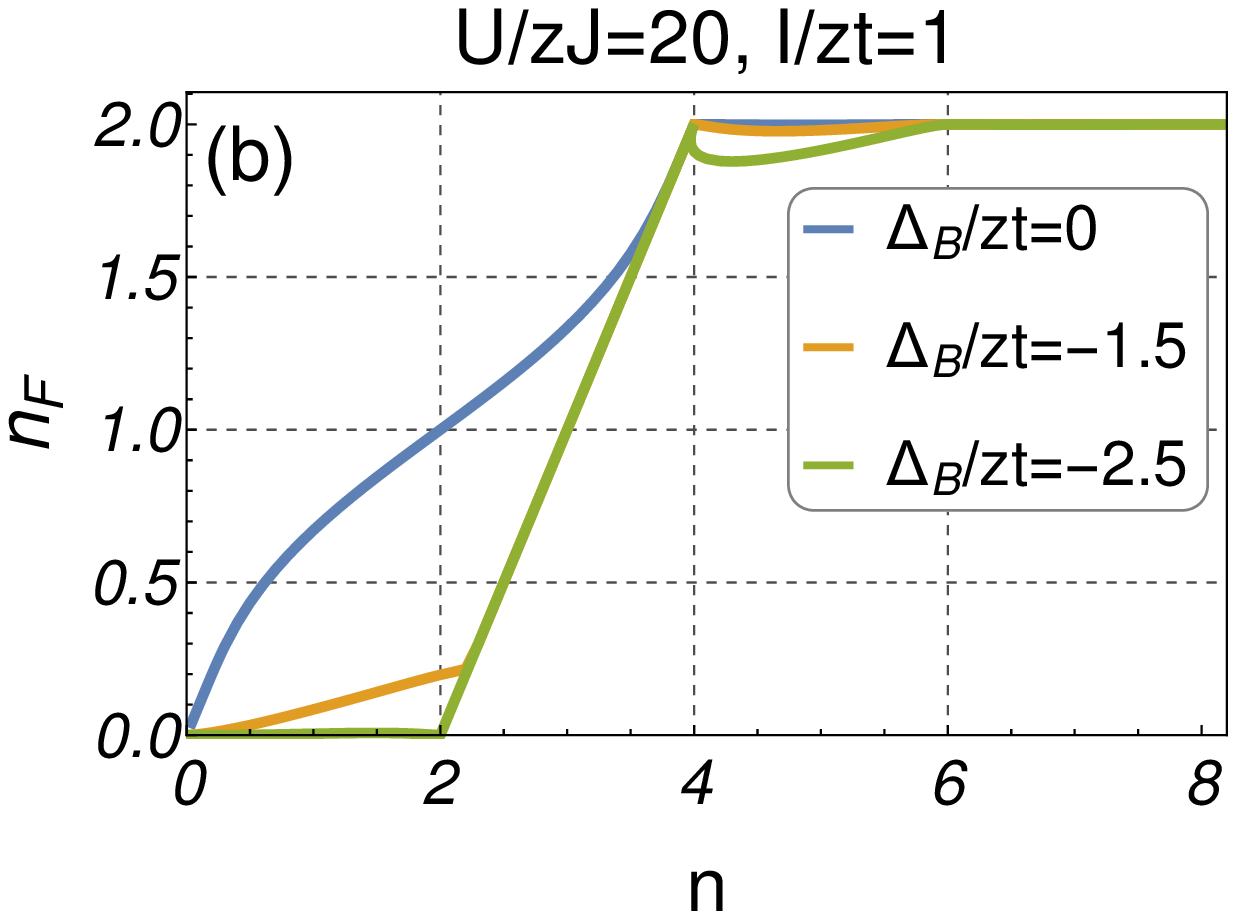}\includegraphics[scale=0.32]{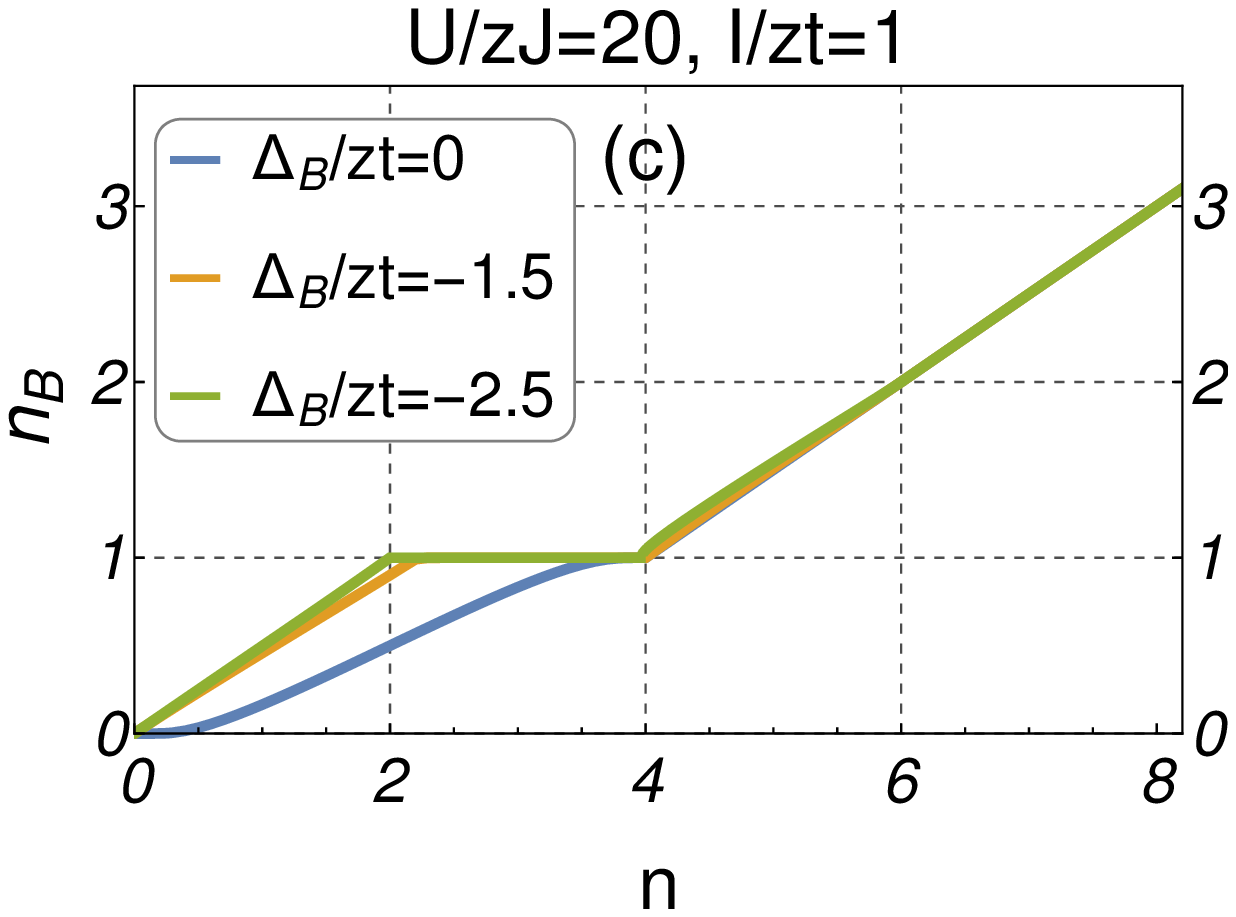}

\includegraphics[scale=0.32]{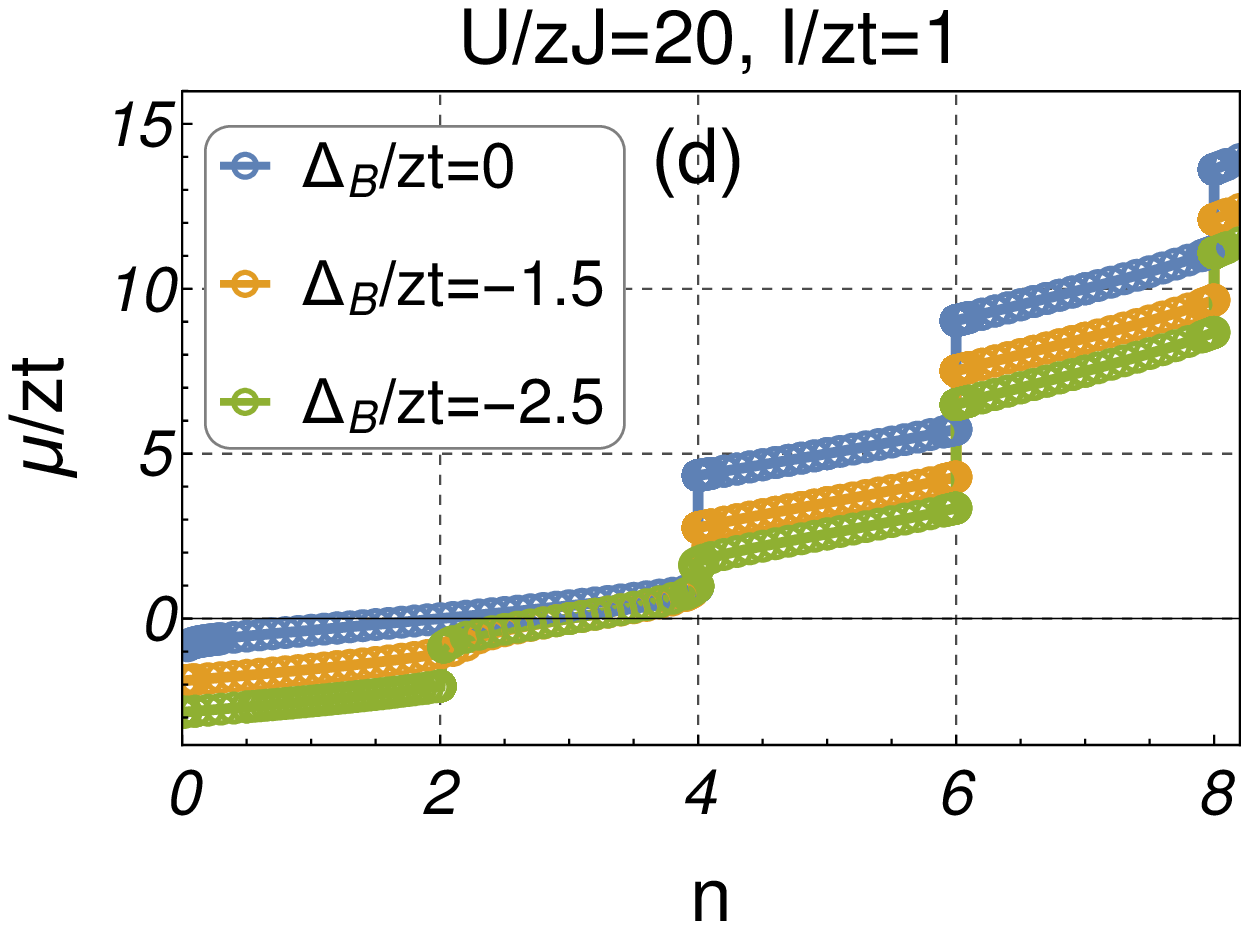}\includegraphics[scale=0.32]{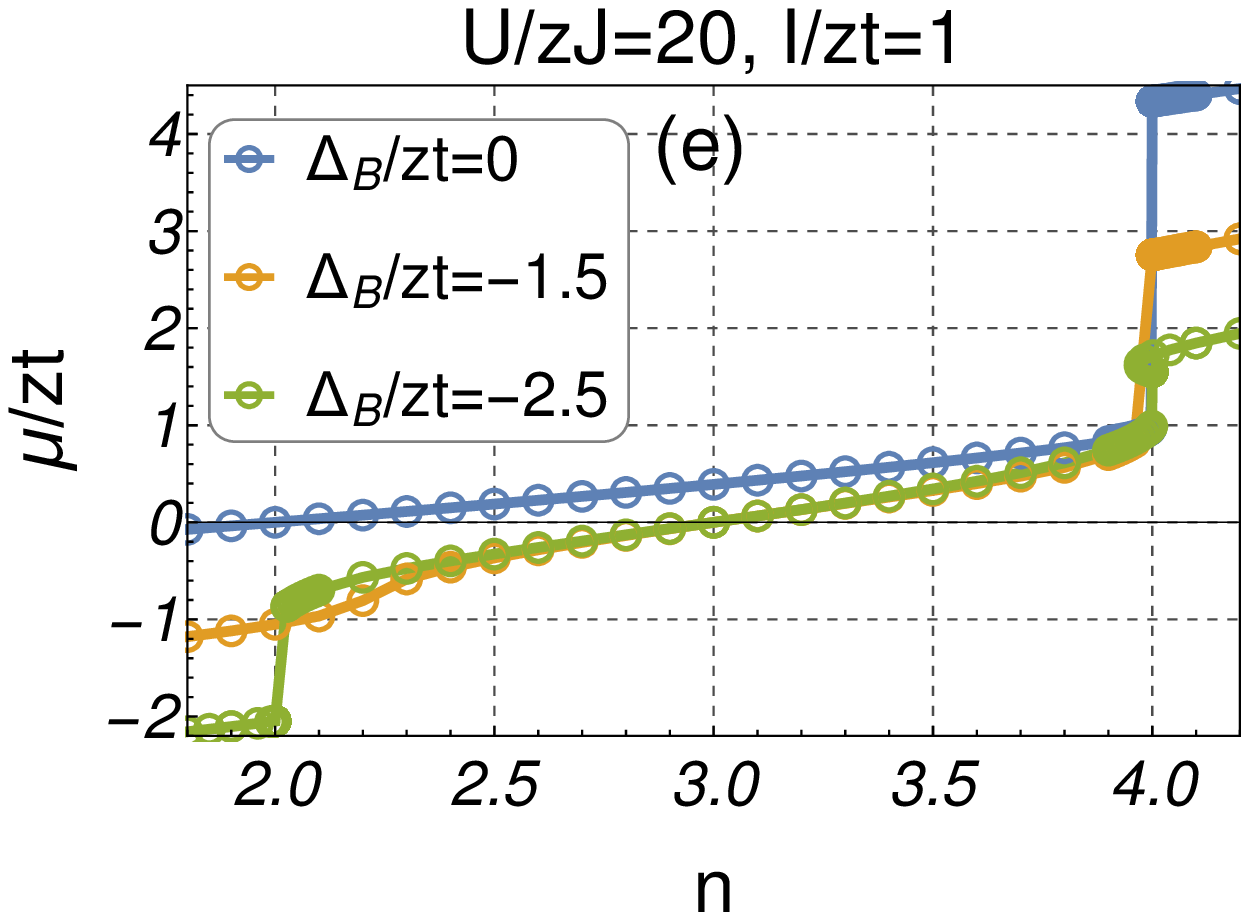}

\caption{(a) Finite temperature mean-field phase diagram of BFHM versus total
particle number $n=2n_{B}+n_{F}$ for different strengths of detuning
parameter $\Delta_{B}$ (see legend). Figures (b), (c), (d) are plots
of $n_{F}$, $n_{B}$, $\mu/zt$ versus $n$, respectively (the data
obtained are evaluated along the critical line from Fig. (a)). Figure
(e) is an enlargement of the vicinity of zero chemical potential from
plot (d). Plots are made assuming that $U/zJ=20$, $I/zt=1$, $J=t/2$.
For comparison, we plot $\Delta_{B}/zt=0$ from Fig. \ref{fig: Delta=00003D0}.
For clarity, the circles are added on the numerical data points in
Figs. (d) and (e). \label{fig: negative detuning}}
\end{figure}

\subsection{Average particle number \label{sub:Average-particle-number}}

During the analysis of the boson-fermion mixture phase diagram in
the following sections, the additional considerations of the average
particle number per site $n$ are made; $n$ is calculated within
the unperturbed part of the action from Eq. (\ref{eq:dzialanie_U_rowne_0-1})
at the phase boundary (it is consistent with the mean-field calculation
of average particle number per site at phase boundary within the operator
approach method, see Appendix \ref{sub: meanfield - lienar response}).
This means that the $0$-th order partition function has the form
$Z_{0}=Z_{0}^{F}Z_{0}^{B}$ where $Z_{0}^{F}=\int\mathcal{D}\left[\bar{c},c\right]e^{-S_{0}^{F}\left[\bar{c,}c\right]/\hbar}$
and $Z_{0}^{B}$ is defined in Eq. (\ref{eq:Z0B}). Therefore $n$
is calculated by using $n=-\partial\ln Z_{0}/\partial\mu$ and we
get
\begin{equation}
n=n_{F}+2n_{B}\;,\label{eq:nFB}
\end{equation}
where $n_{F}$ is the average particle number of fermions for both
spin components
\begin{equation}
n_{F}=2\sum_{\mathbf{k}}\frac{1}{e^{\beta\left(t_{\mathbf{k}}-\mu\right)}+1}\;,\label{eq:nF}
\end{equation}
and $n_{B}$ is an average particle number of bosons 
\begin{equation}
n_{B}=\frac{\sum_{n_{0}=0}^{\infty}n_{0}e^{-\beta E_{n_{0}}}}{\sum_{n_0=0}^{\infty}e^{-\beta E_{n_{0}}}}\;,\label{eq:nB}
\end{equation}
where on-site bosonic energy $E_{n_{0}}$ is defined in Eq. (\ref{eq:E_n0}).
There is also a possibility to obtain Eqs. (\ref{eq:nFB}-\ref{eq:nB})
directly by taking into account Gaussian fluctuations over a saddle
point action $S_{MF}^{eff}$ from Eq. (\ref{eq: effective action 3})
at the phase boundary.

It is also worth adding here that improved approach which includes
the effect of resonant interaction $I$, bosonic hopping $J$ and
fermionic interaction $V$, in the normal phase, can be achieved by
using the self-consistent T-matrix theory \cite{Micnas2014,PhysRevB.76.184507}.

\begin{figure}[th]
\includegraphics[scale=0.65]{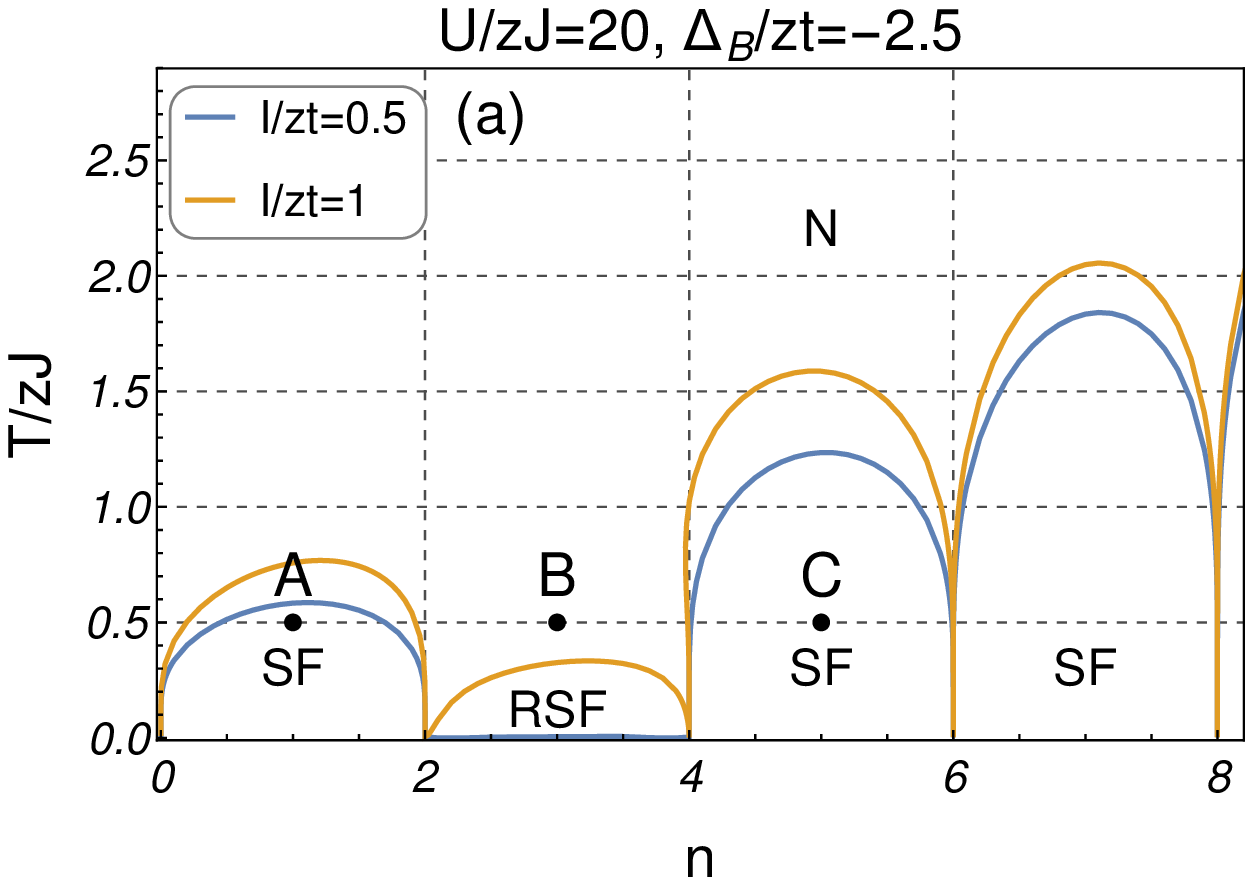}

\includegraphics[scale=0.32]{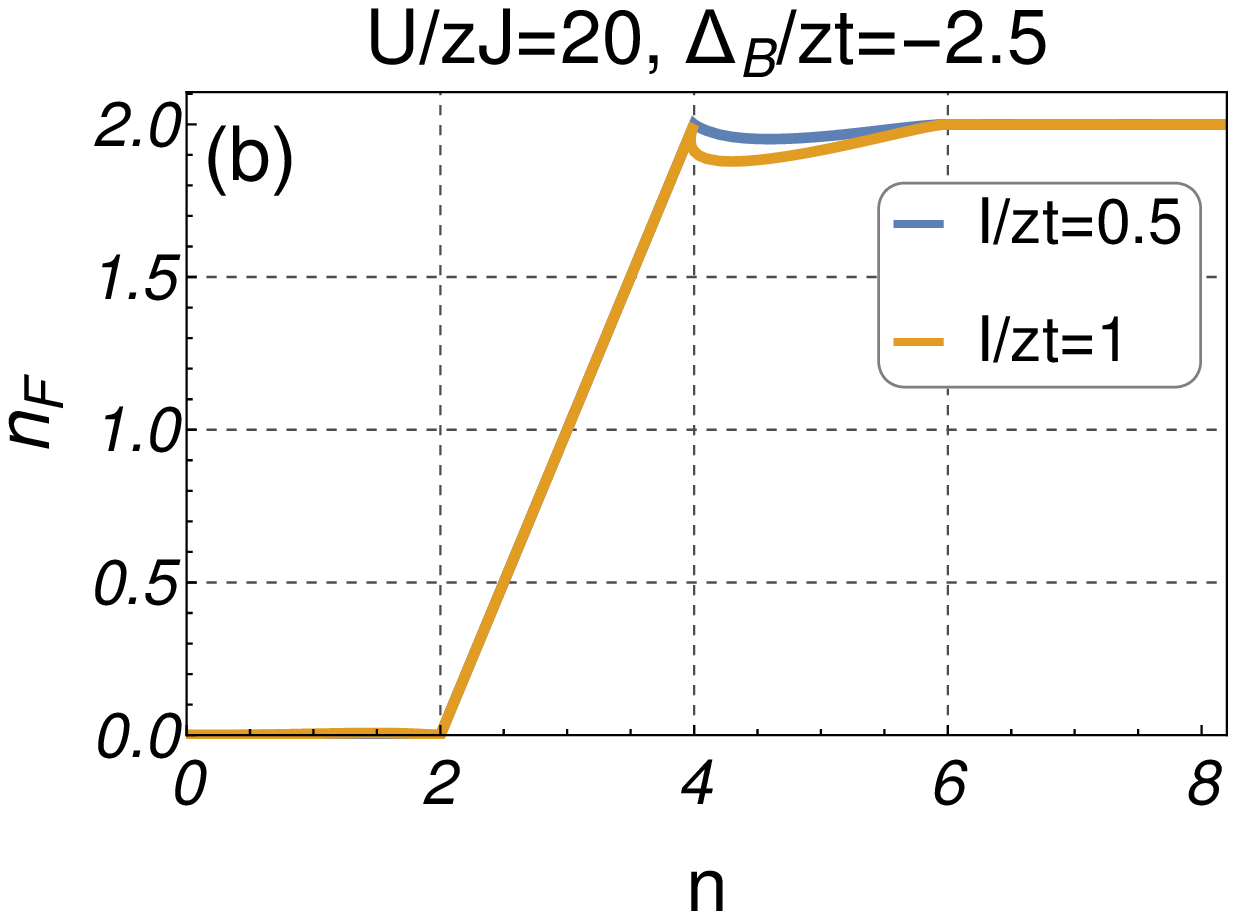}\includegraphics[scale=0.32]{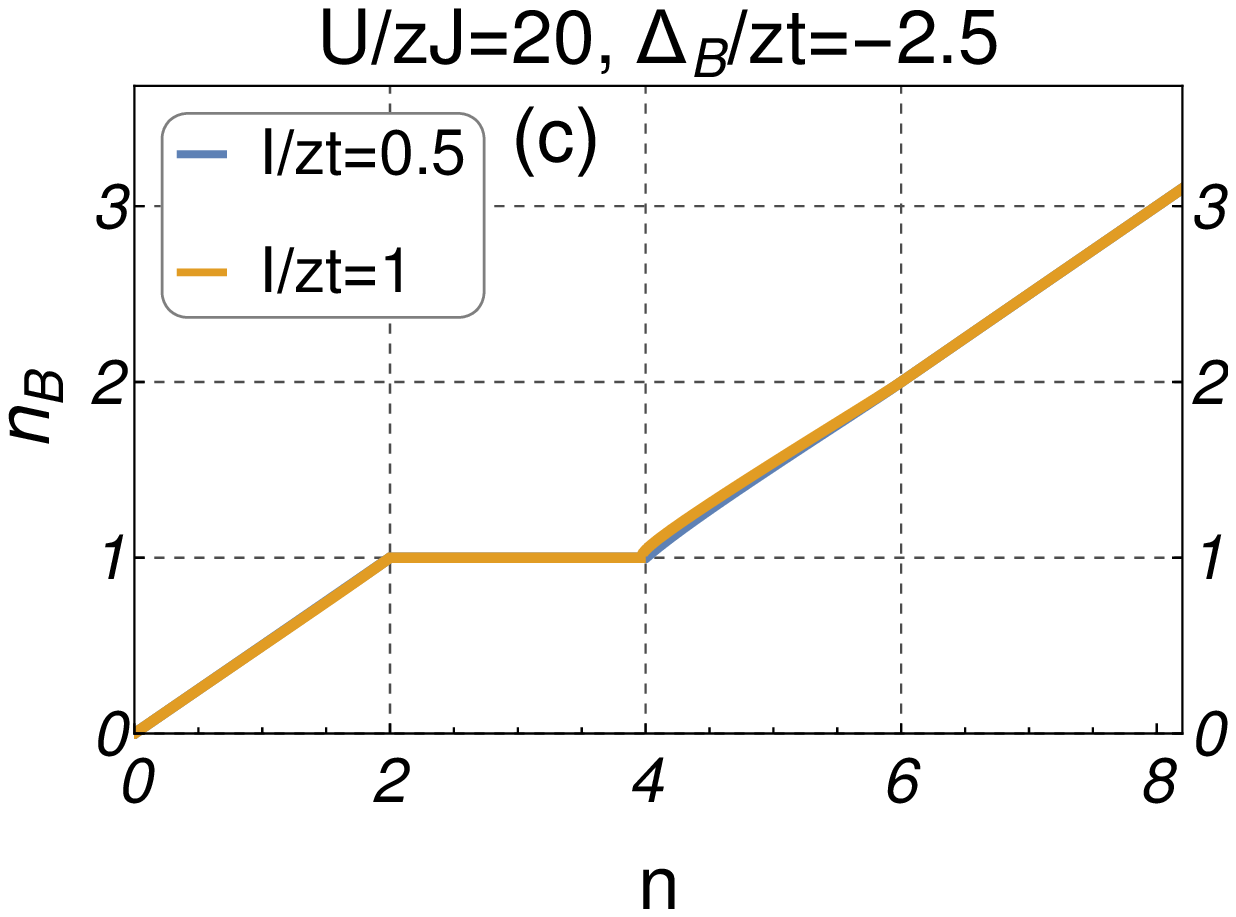}

\includegraphics[scale=0.32]{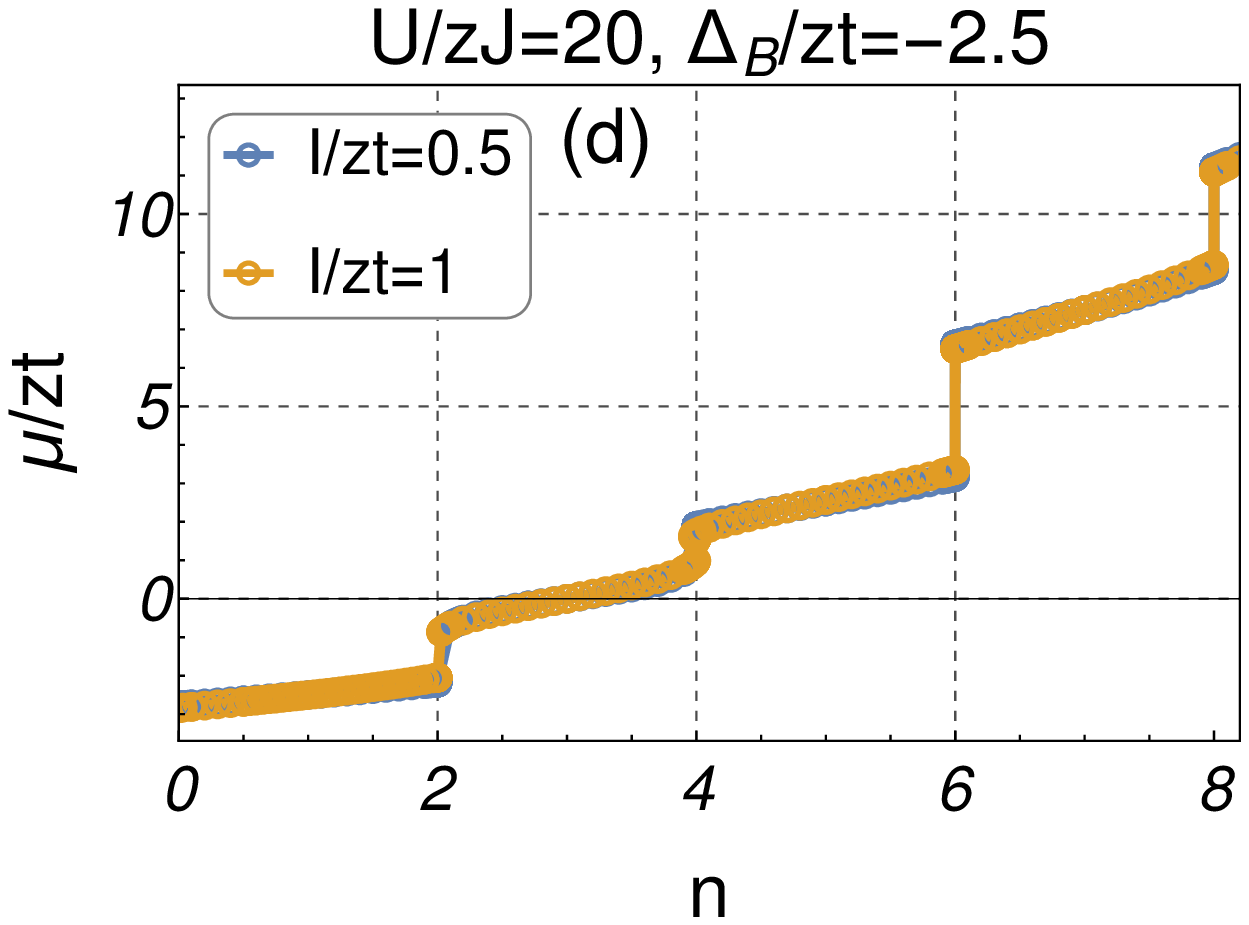}\includegraphics[scale=0.32]{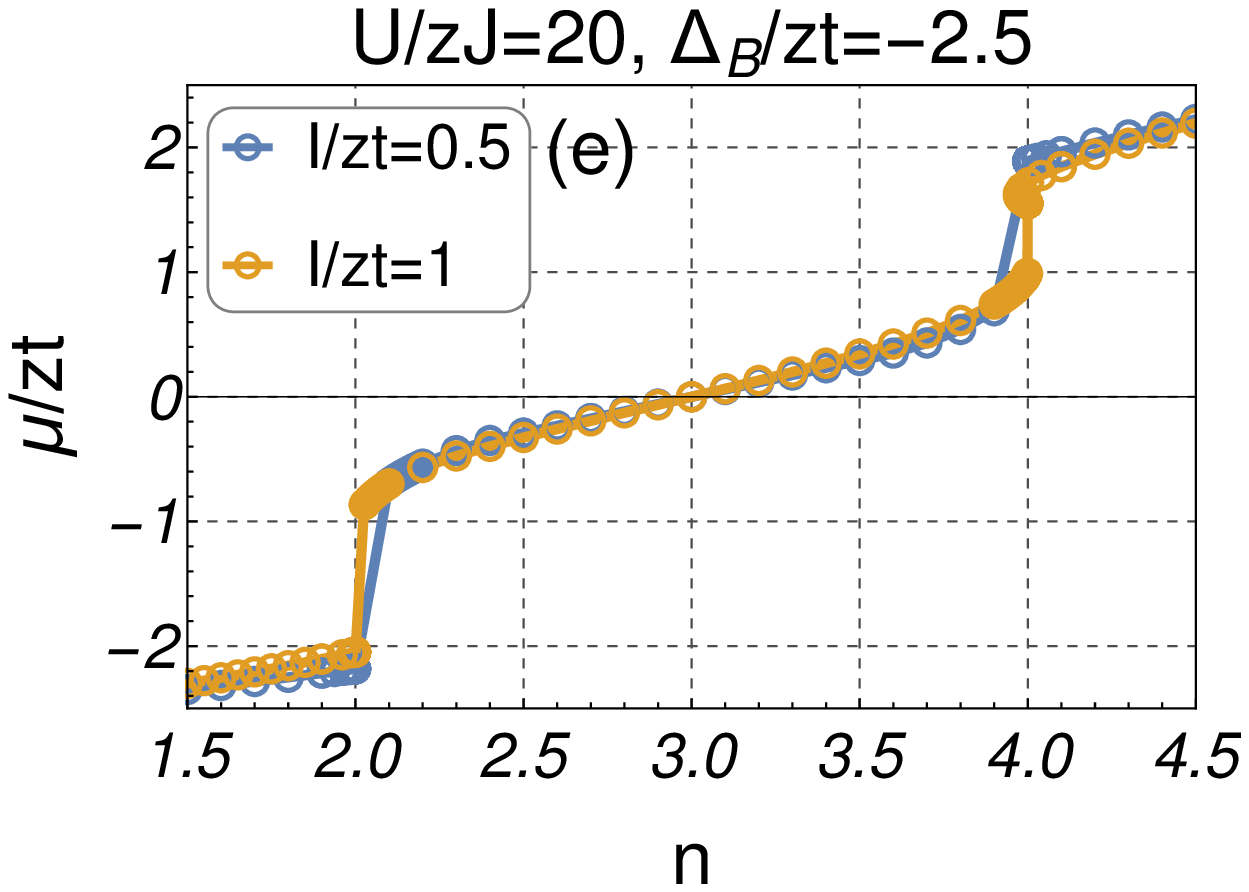}

\caption{(a) Finite temperature phase diagram of BFHM versus total particle
number $n=2n_{B}+n_{F}$ for different strengths of converting interaction
$I/zt$ (see legend). Figures (b), (c), (d) are plots of $n_{F}$,
$n_{B}$, $\mu/zt$ versus $n$, respectively (the data obtained are
evaluated along the critical line from Fig. (a)). Figure (e) is an
enlargement of the vicinity of zero chemical potential from plot (d).
Plots are made assuming that $U/zJ=20$, $\Delta_{B}/zt=-2.5$, $J=t/2$.
For clarity, the circles are added on the numerical data points in
Figs. (d) and (e). Meaning of A, B and C points is given in Sec. \ref{sub:experiment}.}
 \label{fig: I/zt=00003D 1 0.5}
\end{figure}

\begin{figure}[th]
\includegraphics[scale=0.65]{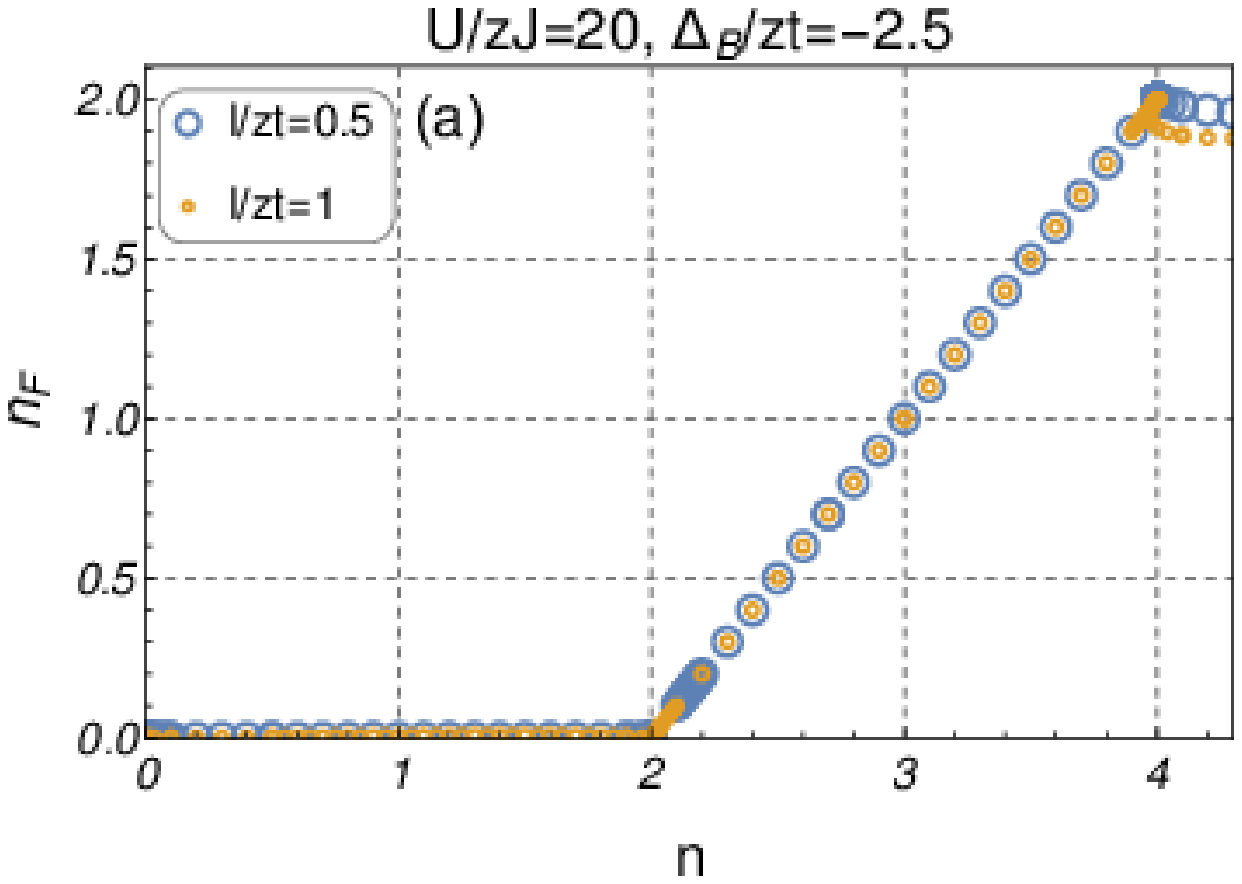}

\includegraphics[scale=0.65]{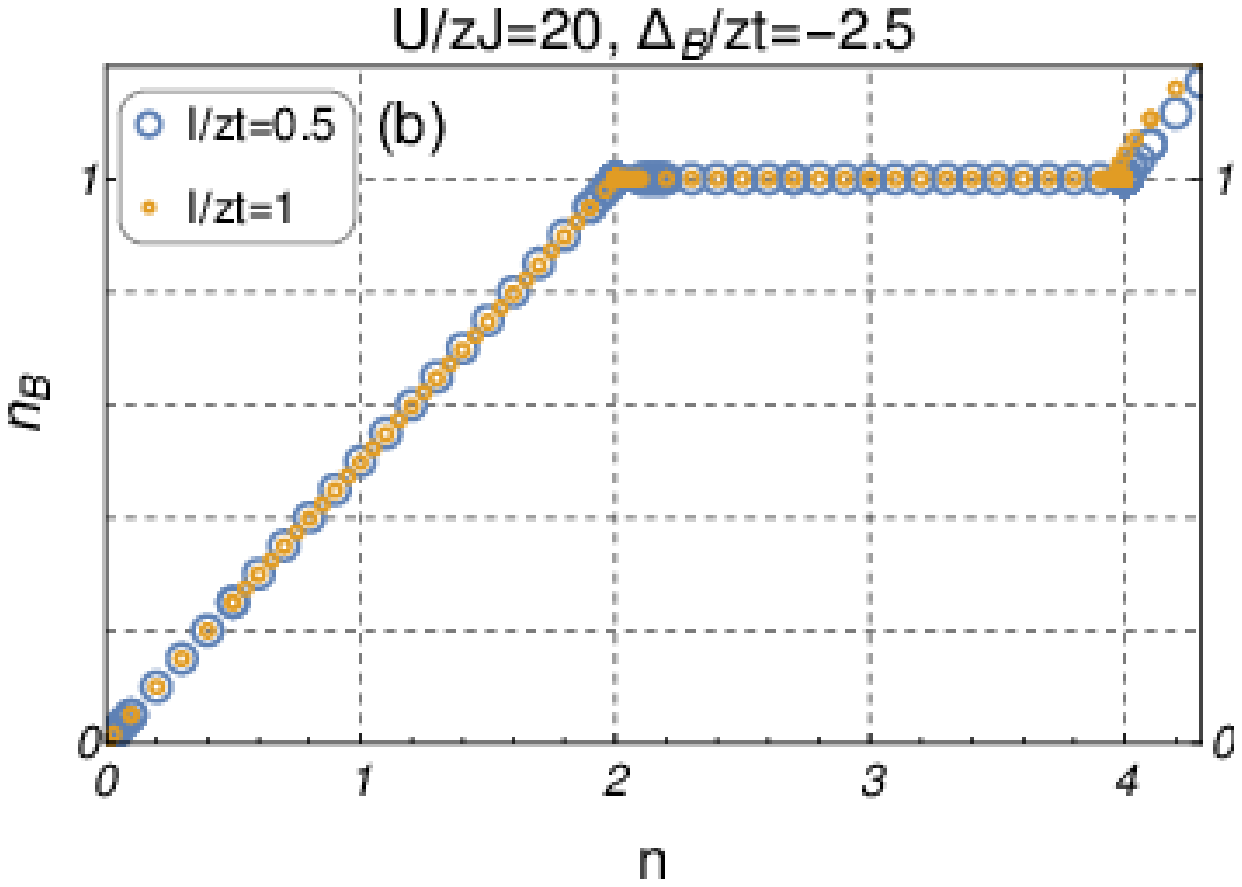}

\caption{Plots a and b are enlargements of the relevant parts of Figs. \ref{fig: I/zt=00003D 1 0.5} b and c, respectively.\label{fig: zoom negative detuning}}
\end{figure}

\begin{figure}[th]
\includegraphics[scale=0.65]{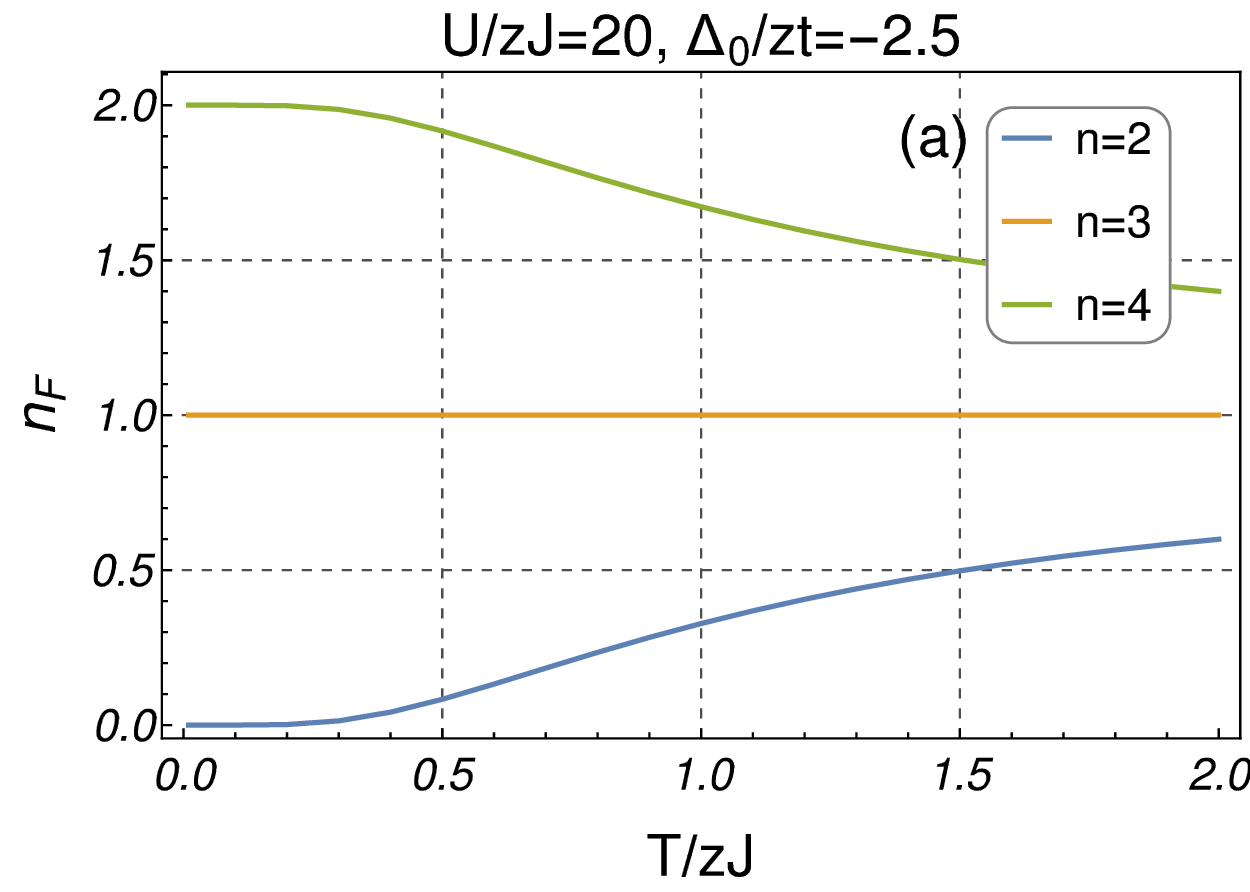}

\includegraphics[scale=0.65]{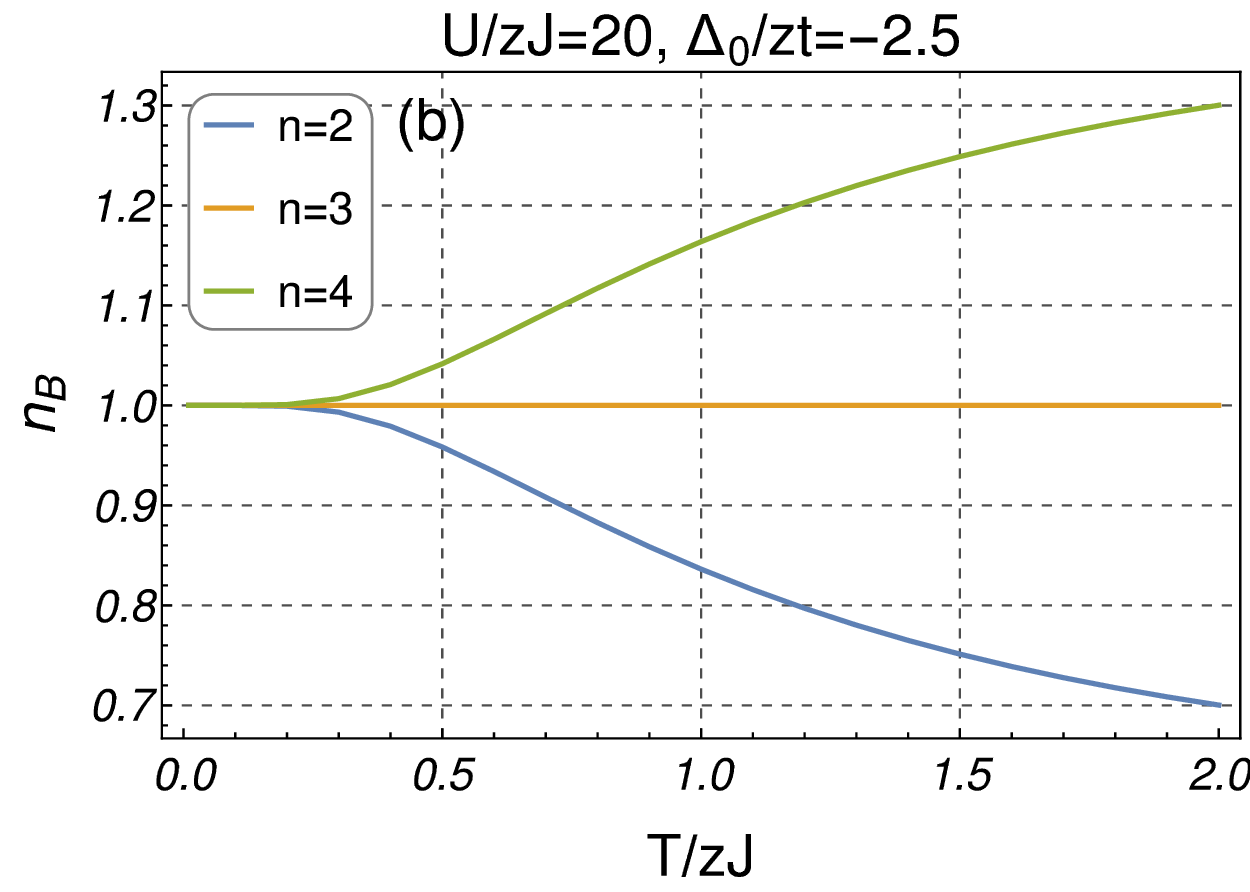}

\caption{Fermionic $n_{F}$ (a) and bosonic $n_{B}$ (b) average particle density
per site with fixed total number of particles $n=n_{F}+2n_{B}$ (see Eqs. (\ref{eq:nFB})-(\ref{eq:nB})). The
other parameters are $U/zJ=20$, $J=t/2$.}
 \label{fig. normal-phase densities}
\end{figure}

\section{Results and discussion \label{sec:Results-and-discussion}}

\subsection{Phase diagram of the BHM}

In order to clarify further discussion, we shortly review the finite
temperature phase diagram of the standard BHM in terms of reduced
critical temperature $T_{c}/zJ$ versus average concentration of bosons
per site $n_{B}$. 

Using previously defined bosonic annihilation and creation operators
$b_{i}$ and $b_{i}^{\dagger}$, BHM Hamitonian has the form $H_{BHM}=-\sum_{ij}\left(J_{ij}+\mu\delta_{ij}\right)b_{i}^{\dagger}b_{j}+U\sum_{i}b_{i}^{\dagger}b_{i}^{\dagger}b_{i}b_{i}$.
The phase diagram comprising SF, bosonic Mott insulator (BMI) and
normal (N) phases is well-known \cite{Sheshadri:2007wc,Trotzky:2010ti,Sajna2015}
and in the mean-field approximation the critical line is given by
$\epsilon_{\mathbf{0}}-\left[G^{1,c}\left(i\nu_{n}=0\right)\right]^{-1}=0$.
In Fig. \ref{fig: pure BHM}, we plot critical temperature $T_{c}/zJ$
dependence on the average density of bosons per site $n_{B}$ for
the critical boundary in BHM. BMI for different integer values of
$n_{B}$ are located only between lobes at zero temperatures which
are indicated in Fig. \ref{fig: pure BHM} by black arrows (at finite
temperatures there is no true insulating state \cite{Gerbier:2007to}).
Here and in the following subsection we choose $U/Jz=20$ to analyze
strong interaction limit of bosonic particles.

\subsection{Density phase diagram of BFHM model \label{sub:Phase-diagram-of-BFM}}

We are interested in the density phase diagram of BFHM in the limit
$J\ll U$ and $V=0$ (as was mentioned in Sec. \ref{sec: phase diagraaam}). The critical boundary line at finite temperatures is obtained
from Eq. (\ref{eq:critical-line-equation}). In the following subsections
\ref{sub:Zero-detuning}, \ref{sub:Negative-detuning}, \ref{sub:Positive-detuning},
the phase diagram of BFHM is analyzed in three different regimes of
parameter $\Delta_{B}$ which controls mutual position of fermionic
and bosonic band, namely: (a) $\Delta_{B}/zt=0$, (b) $\Delta_{B}/zt>0$,
(c) $\Delta_{B}/zt<0$. In particular, the value of parameter $\Delta_{B}$
is directly related to the position of the bottom of bosonic band
with respect to that of fermionic one. It is clear from considering
BFHM Hamiltonian from Eq. (\ref{eq:BFM-hamiltonian}) and from relation $J=t/2$
  which corresponds to assumption that one molecule is made of two fermionic particles. The bottom of the boson band is located
at the center of the fermion band at $\Delta_{B}/zt=0.25$ and it
starts to appear below the fermionic band for $\Delta_{B}/zt<-0.75$
and above for $\Delta_{B}/zt>1.25$.

\subsection{Zero detuning ($\Delta_{B}=0$) \label{sub:Zero-detuning}}

In Fig. \ref{fig: Delta=00003D0}, we show the finite temperature
phase diagram for BFHM with zero detuning $\Delta_{B}/zt=0$, finite
bosonic interaction strength $U/zJ=20$ and converting interaction
$I/zt=1$. These results explicitly show that if $\Delta_{B}/zt=0$
parameter is close to $\Delta_{B}/zt=0.25$ value (i.e. bottom of
bosonic band is close to the middle of fermionic one), the critical
line assumes a regular lobe structure similarly like in the phase
diagram of standard BHM (see Fig. \ref{fig: pure BHM}). However,
in BFHM case the lowest lobe is relatively wider than the others (i.e.
$n\in\left(0,4\right)$ instead of width $2$ in $n$ units in comparison
to pure BHM case, see. Fig. \ref{fig: pure BHM}). This widening is
related to the gradual filling up of the fermionic band with increasing
value of total particles $n$ (see, Fig. \ref{fig: Delta=00003D0}
b). Indeed, chemical potential gradually crosses the fermionic band
which is clearly visible in Fig. \ref{fig: Delta=00003D0} d and e,
i.e. $\mu/zt$ appear at the bottom of fermionic band ($\mu=-zt$)
at $n=0$ and ending at the top of fermionic band ($\mu=zt$) for
$n=4$.

It is interesting to notice here that  in comparison to the BHM case (Fig. \ref{fig: pure BHM}), there is an enhancement of the superfluid critical temperature when $I\neq 0$. Starting from second lobe, this enhancement can be simple accounted for  the pairing mechanism of fermionic holes. This is confirmed by the slight deviations of fermionic density from a band insulator regime ($n_F=2$) for $n>4$ (see Fig. \ref{fig: Delta=00003D0} b and its corresponding enlargement in Fig. \ref{fig: supplementtoFig2}).

The above picture is dramatically changed when detuning starts to
deviate from zero value. It will be discussed below.

\subsection{Positive detuning ($\Delta_{B}>0$) \label{sub:Positive-detuning}}

With increasing value of $\Delta_{B}/zt$ parameter, the bottom of
bosonic band is above the fermionic one for $\Delta_{B}/zt>1.25$.
This should result in increasing fermionic density at the expense
of bosonic one at low $n$ which indeed is clearly visible in Fig.
\ref{fig: positive detuning} a-c. In particular, with increasing
$\Delta_{B}/zt$, the lower part of the first lobe gradually diminishes
and the first lobe-like structure appears for $n\in\left(2,4\right)$
(see, Fig. \ref{fig: positive detuning} with $\Delta_{B}/zt=1.5$).
Such a situation is also confirmed by analysis of the chemical potential
$\mu/zt$ (see, Fig. \ref{fig: positive detuning} d and e) which
shows that its value starts to appear only in region of fermionic
band for $n\in\left(0,\,2\right)$ and for higher values of $\Delta_{B}/zt$
for which the bosonic density is very low (it should be compared to
the situation with $\Delta_{B}/zt=0$ in which $\mu\in\left\langle -zt,\, zt\right\rangle $
for $n\in\left(0,\,4\right)$, Fig. \ref{fig: positive detuning}
d and e).

\subsection{Negative detuning ($\Delta_{B}<0$) \label{sub:Negative-detuning}}

The situation is even more interesting for negative detuning for which
the bottom of bosonic band is below the fermionic one for $\Delta_{B}/zt<-0.75$.
Intuitively, when the number of particles $n$ is increased, at first
the bosonic band should start to fill up. This intuition fully agrees
with our simulation presented in Fig. \ref{fig: negative detuning}
for $n_{B}$ and $n_{F}$ versus $n$ and is clearly observed in the
regime of relatively high negative values of $\Delta_{B}/zt=-2.5$.
However, in comparison to the reference case at $\Delta_{B}/zt=0$
the situation here is more complex, the critical line at $\Delta_{B}/zt=-2.5$
for $n\in\left\langle 0,\,4\right\rangle $ range decays into two
lobes (see, Fig. \ref{fig: negative detuning} a). The first lobe
at $n\in\left\langle 0,\,2\right\rangle $ contains the SF phase with
gradually increasing average number of bosonic particles $n_{B}$
(Fig. \ref{fig: negative detuning} c) and the second lobe at $n\in\left\langle 2,\,4\right\rangle $
is characterized by the almost integer bosonic density $n_{B}$ (here close to one)  i.e. it has the BMI character for bosonic particles (the bosonic density deviates from the integer number with order less than $10^{-6}$)
(Fig. \ref{fig: negative detuning} c). Moreover, the fermionic part
for $n\in\left\langle 2,\,4\right\rangle $ gradually changes its
density from $n_{F}=0$ to $n_{F}=2$ with increasing value of $n$.
We also clearly see, that the phase is characterized by location of
chemical potential inside the fermionic band, pointing out that the
system is at the Feshbach resonance (see, Fig. \ref{fig: negative detuning}
d and e). Further, we argue that this superfluid phase with 
number of bosons close to integer value arises purely from the resonant mechanism and for
simplicity we denote it as resonant superfluid (RSF) phase.

To show the resonant character of RSF we check the sensitivity of
this phase by tuning the amplitude of converting interaction in $S_{0}^{FB}$
from Eq. (\ref{eq:dzialanie_U_rowne_0-1}). Namely, in Fig. \ref{fig: I/zt=00003D 1 0.5} 
we plot the phase diagram for different values of $I/zt$. This phase
diagram shows that the RSF phase is highly suppressed at finite temperatures
and it almost disappear for $I/zt=0.5$. Therefore one can conclude
that RSF phase originates from the Feshbach-like correlations. 

Moreover, it is worth adding here, that fermionic $n_F$ and bosonic $n_B$ densities are almost intact with respect to the change of $I/zt$ in RSF phase (see Fig. \ref{fig: I/zt=00003D 1 0.5}  b, c and \ref{fig: zoom negative detuning}). However, as expected we observe, that there is a slight change of these densities not visible in the presented density plots (the order of this change is less than $10^{-6}$).

We also checked the vicinity of the RSF region by analyzing normal
phase above the critical temperature in terms of $n_{F}$ and $n_{B}$
densities at a constant $n$ (see, Fig. \ref{fig. normal-phase densities}).
These densities correspond to the $I/zt=0$ regime at this level of
approximation (see Sec. \ref{sub:Average-particle-number} and Eqs.
(\ref{eq:nFB}-\ref{eq:nB}). From Fig. \ref{fig. normal-phase densities},
we observe that in the $T\rightarrow0$ limit, $n_{B}$ is pinned
to the integer value equal to one while $n_{F}$ gradually increases
for the corresponding total particle density $n=2,\,3,\,4$. This
observation is consistent with the conclusion about RSF phase drawn
in the previous paragraph.

It is also worth adding here, that the above picture of BFHM phase
diagram is also consistent with the work \cite{Micnas2003} which
considered the hard-core limit of bosonic particles without bosonic
hopping ($J=0$). It should not be surprising because our theory properly
recover this limit at the mean-field level (see, Eq. \ref{eq: order paramter - HC}).
However, RSF phase with the number of bosons close to one is a novel
behavior which appears beyond the hard-core limit.

Moreover, when the system is beyond the Feshbach resonance for $\Delta_{B}/zt=-2.5$
(i.e. chemical potential is below or above fermionic band) there is
another interesting feature observed in Fig. \ref{fig: I/zt=00003D 1 0.5}.
Namely, the SF phase is favored for $n\in\left(0,\,2\right)$ and
$n>4$, but it is important to point out here that the mechanism behind
it is quite different. In the $n\in\left(0,\,2\right)$ range SF is
enhanced through paring of fermionic particles (BCS like character),
but in the $n>4$ range paring mechanism is through fermionic holes.
It is indicated by the corresponding low magnitude enhancement (for
$n\in\left(0,\,2\right)$) or reduction (for $n>4$) of the fermionic
density part in numerical data.

At the end of this section, we would like to also add that for higher
values of negative detuning $\Delta_{B}/zt$, the general behavior
of phase boundary is similar to that discussed above. Namely, higher
negative values of $\Delta_{B}/zt$ shift of chemical potential also
to higher negative values causing that Feshbach resonance region around
$\mu/zt\in\left\langle -1,\,1\right\rangle $ appears for higher densities.
Then, depending on the $\Delta_{B}/zt$ value, a situation like that
in the former cases appears, i.e. (1) the widening of one of the lobes
like for $\Delta_{B}/zt=0$ (see, Fig. \ref{fig: positive detuning})
or (2) the emergence of RSF mixture like for $\Delta_{B}/zt=-2.5$
(see, Fig. \ref{fig: negative detuning}). In particular, up to $\Delta_{B}/zt=-10$
with the same BFHM Hamiltonian parameters as before, we numerically
check, that the first situation (1) appears for $\Delta_{B}/zt=-5$
and $\Delta_{B}/zt=-10$ and the second one (2) appears for $\Delta_{B}/zt=-7.5$
(here RSF phase emerge for $n\in\left(4,\,6\right)$).

It would be also interesting in further investigations beyond mean-field
approximation, to include the effects of pairing fluctuations into
theory which should imply lowering of superfluid critical temperature.
Then the temperature obtained in this work will correspond to the
appearance of the pseudogap regime for fermionic particles \cite{Micnas2014,PhysRevB.76.184507,2002PhRvB..66j4516M}.

\subsection{RSF phase in the time of flight type experiment} \label{sub:experiment}

Time of flight (TOF) type spectroscopy is one of the most powerful
methods of measurements in the state of art of current experimental
setups in ultracold atoms. Within the optical lattice systems, it
has been widely used for e.g. bosons \cite{Kennedy_2015,PhysRevLett.101.155303,PhysRevLett.98.080404,Gerbier:828201,Greiner:2002wt},
fermions \cite{Rom_2006,Chin_2006} or boson-fermion mixtures \cite{PhysRevLett.102.030408,PhysRevLett.96.180402}.
In particular, it is relatively simple to probe coherence via momentum
distribution encoded in freely expanding cloud. As an example, it
has been previously used to detect SF-BMI quantum phase transition
in the bosonic Rb atoms \cite{Greiner:2002wt} or resonant superfluidity
in the fermionic Li atoms \cite{Chin_2006}. In a realistic experiment,
the enhancement of coherence is observed as the appearance of peaks
in the time of flight pattern \cite{Kennedy_2015,Chin_2006,Gerbier:828201,Greiner:2002wt}.

We suggest that the footprint of the RSF phase can be tested by preparing
ultracold fermionic gas at the Feshbach resonance with negative detuning
of $\Delta_{B}$ parameter. The detuning should be about two and half
times greater than the width of the fermionic band. Then repeating
the experiment with increasing number of fermions which simulate BFHM
(which is close to the ground state), one should observe a lowering
of coherence at $n\in\left\langle 2,\,4\right\rangle $ densities.
It can be deduced from the phase diagram in Fig. \ref{fig: I/zt=00003D 1 0.5}
where in the range $n\in\left(0,\,2\right)$ and $n>4$, SF phase
has a higher critical temperature than in the $n\in(2,\,4)$ region. 

For instance, let's assume that the atomic gas is prepared at similar
temperatures for different particle numbers which are represented
by points A, B and C in Figs. \ref{fig: positive detuning} a and
\ref{fig: I/zt=00003D 1 0.5} a. Furthermore, let's assume that in
each of these phases represented by points A, B and C, TOF experiment
is performed. Then, it can be concluded that for the situation with
positive detuning as in Fig. \ref{fig: positive detuning} a, the coherence
of bosonic particles should be an increasing function of $n$ at corresponding
points A, B and C, because of the deeper penetration of the system
into SF phase for A, B and C, respectively. However, this situation
should be quite different for negative detuning of $\Delta_{B}/zt$.
As shown in Fig. \ref{fig: I/zt=00003D 1 0.5} a, point B in comparison
to point A and C is located beyond SF phase, which means that TOF
pattern does not exhibit the behavior characteristic of SF phase \cite{Trotzky:2010ti}.
Therefore, for negative detuning, one should observe non-monotonous
behavior of coherence peaks which can be read off from TOF patterns
for the corresponding points A, B and C. Moreover, increasing strength
of Fershbach interaction $I/zt$ should result in gradual disappearance
of this non-monotonous behavior at point B (see Fig. \ref{fig: I/zt=00003D 1 0.5}
a). Consequently, such coherence dependence which can be observed
in experiment, could be accounted for by the appearance of RSF phase
in the investigated system.

\section{Summary}

In this work, we investigated the limit of strongly correlated Feshbach
molecules at finite temperatures in a three dimensional lattice. We
show, that for negative detuning $\Delta_{B}/zt$ and at least for
weak strength of converting interaction $I/zt$, a resonant superfluid
phase (RSF) appears which is characterized by an arbitrary number
of fermions per site (i.e. fermionic concentration between 0 and 2)
and an integer number of bosonic atoms. This happens when fermions
are in the Feshbach resonance. We show that this resonant character
of RSF phase is unstable toward weakening converting interaction $I/zt$.
In the situation when the fermions are beyond resonance the superfluid
phase is strengthened. We explain that this enhancement is caused
by hole pairing mechanism for higher densities, while for lower densities
it is standard fermionic particle paring mechanism which corresponds
to that known in the BCS theory.

Moreover, we have also discussed the experimental protocol in which
footprint of RSF phase can appear in TOF type experiment. Namely,
the footprint of the RSF phase could be simply observed as a non-monotonous
behavior of coherence peaks from time of flight pattern when the number
of fermions is increased.

In future investigation, it will be also interesting to study the
system's behavior from the point of view of tuning the parameter $\Delta_{B}$
at fixed total $n$. Especially interesting analysis would be for
the total density equal to two ($n=2$) in which two different peculiar
regimes should appear depending on the $\Delta_{B}$ and $U$ amplitude.
Namely, tuning the system from positive $\Delta_{B}>0$ to negative
value $\Delta_{B}<0$, should result in transition from fermionic
band insulator ($n_{F}=2$, $n_{B}=0$) to SF phase and from SF to
bosonic Mott insulator ($n_{F}=0$, $n_{B}=1$). We left this problem
for future studies in which careful analysis of the BFHM ground state
is also required.
\begin{acknowledgments}
We would like to thank to Prof. T. K. Kope\'{c} for useful discussions
on the early stage of the presented work. We are also grateful to
Dr T. P. Polak for careful reading of the manuscript.
\end{acknowledgments}

\section*{Appendix}

\subsection{Local Green function \label{sub:Local-Green's-function}}

On-site single particle green function, defined as $\frac{1}{\hbar}G^{1,c}\left(\tau-\tau'\right)=-\left\langle \bar{\psi}_{i}(\tau)\psi_{i}(\tau')\right\rangle _{0}^{B}$
is given by

\begin{equation}
\frac{1}{\hbar}G^{1,c}\left(i\nu_{n}\right)=\frac{1}{Z_{0}}\sum_{n_{0}=0}^{\infty}(n_{0}+1)\frac{e^{-\beta E_{n_{0}+1}}-e^{-\beta E_{n_{0}}}}{E_{n_{0}+1}-E_{n_{0}}-i\hbar\nu_{n}}\;,\label{eq: G0}
\end{equation}
where

\begin{equation}
E_{n_{0}}=-\mu^{*}n_{0}+Un_{0}(n_{0}-1)/2\;,\label{eq:E_n0}
\end{equation}
\begin{equation}
Z_{0}=\sum_{n_{0}=0}^{\infty}e^{-\beta E_{n_{0}}}\;.
\end{equation}

\begin{widetext}
\subsection{Generating functional in the BFHM} \label{sub: generating functional}
The generating function of statistical sum from Eq. (\ref{eq:global-partition-function}) has
the form
\begin{equation}
Z\left[\bar{\gamma,}\gamma\right]=\int\mathcal{D}\left[\bar{c},c,\bar{b,}b\right]e^{\sum_{ij}\int_{0}^{\hbar\beta}d\tau J_{ij}\bar{b}_{i}(\tau)b_{j}(\tau)-S_{0}^{F}\left[\bar{c,}c\right]-S_{0}^{B}\left[\bar{b,}b\right]-S_{0}^{FB}\left[\bar{b,}b,\bar{c,}c\right]+\sum_{i}\int_{0}^{\beta}d\tau\left(\bar{\gamma}_{i}(\tau)b_{i}(\tau)+c.c.\right)},\label{eq:generarting-global-partition-function}
\end{equation}
where $\gamma_{i}(\tau)$, $\bar{\gamma}_{i}(\tau)$ are external sources. It can be rewritten to the form
\begin{equation}
Z\left[\bar{\gamma,}\gamma\right]=\int\mathcal{D}\left[\bar{c},c,\bar{b,}b\right]e^{\sum_{ij}\int_{0}^{\hbar\beta}d\tau J_{ij}\bar{b}_{i}(\tau)b_{j}(\tau)-S_{0}^{F}\left[\bar{c,}c\right]-S_{0}^{B}\left[\bar{b,}b\right]-\sum_{i}\int_{0}^{\beta}d\tau\left\{ \left[-\bar{\psi}_{i}(\tau)+I\bar{c}_{i\uparrow}(\tau)\bar{c}_{i\downarrow}(\tau)-\bar{\gamma}_{i}(\tau)\right]b_{i}(\tau)+c.c.\right\} }
\end{equation}
After first HS of bosonic fields $b_{i}(\tau)$, $\bar{b}_{i}(\tau)$  (see also Eq. (\ref{eq: 1 HS})), one has
\begin{eqnarray}
 &  & Z\left[\bar{\gamma,}\gamma\right]=Z_{0}^{B}\det\left[\mathbf{J}^{-1}\right]\int\mathcal{D}\left[\bar{c},c,\bar{\psi},\psi\right]\nonumber \\
 &  & \times e^{-\frac{1}{\hbar}\sum_{ij}\int_{0}^{\hbar\beta}d\tau J_{ij}^{-1}\bar{\psi}_{i}(\tau)\psi_{j}(\tau)-\frac{1}{\hbar}\sum_{i}\int_{0}^{\hbar\beta}d\tau\left(\left[-\bar{\psi}_{i}(\tau)+I\bar{c}_{i\uparrow}(\tau)\bar{c}_{i\downarrow}(\tau)-\bar{\gamma}_{i}(\tau)\right]b_{i}(\tau)+c.c.\right)}\nonumber \\
 &  & \times e^{-S_{0}^{F}\left[\bar{c,}c\right]-S_{0}^{B}\left[\bar{b,}b\right]-S_{0}^{FB}\left[\bar{b,}b,\bar{c,}c\right]}.
\end{eqnarray}
Next, shifting $\psi_{i}(\tau)\rightarrow\psi_{i}(\tau)-\gamma_{i}(\tau)+Ic_{i\downarrow}(\tau)c_{i\uparrow}(\tau)$,
$\bar{\psi}_{i}(\tau)\rightarrow\bar{\psi}_{i}(\tau)-\bar{\gamma}_{i}(\tau)+I\bar{c}_{i\uparrow}(\tau)\bar{c}_{i\downarrow}(\tau)$,
we obtain
\begin{eqnarray}
 &  & Z=Z_{0}^{B}\det\left[\mathbf{J}^{-1}\right]\int\mathcal{D}\left[\bar{c},c,\bar{\psi},\psi\right]\nonumber\\
 &  & \times e^{-\frac{1}{\hbar}\sum_{ij}\int_{0}^{\hbar\beta}d\tau J_{ij}^{-1}\left[\bar{\psi}_{i}(\tau)+I\bar{c}_{i\uparrow}(\tau)\bar{c}_{i\downarrow}(\tau)-\bar{\gamma}_{i}(\tau)\right]\left[\psi_{j}(\tau)+Ic_{j\downarrow}(\tau)c_{j\uparrow}(\tau)-\gamma_{i}(\tau)\right]-W_{1}\left[\bar{\psi,}\psi\right]}\nonumber\\
 &  & \times e^{-S_{0}^{F}\left[\bar{c,}c\right]}.
\end{eqnarray}
Finally, taking second HS (see also Eq. (\ref{eq: 2 HS}))
\begin{eqnarray}
 &  & -\sum_{ij}\int_{0}^{\hbar\beta}d\tau\left[\bar{\psi}_{i}(\tau)+I\bar{c}_{i\uparrow}(\tau)\bar{c}_{i\downarrow}(\tau)-\bar{\gamma}_{i}(\tau)\right]\nonumber \\
 &  & \times J_{ij}^{-1}\left[\psi_{j}(\tau)+Ic_{j\downarrow}(\tau)c_{j\uparrow}(\tau)-\gamma_{i}(\tau)\right]\nonumber \\
 &  & \rightarrow\sum_{ij}\int_{0}^{\hbar\beta}d\tau J_{ij}\bar{\phi}_{i}(\tau)\phi_{j}(\tau)\nonumber \\
 &  & -\left\{ \sum_{i}\int_{0}^{\hbar\beta}d\tau\bar{\phi}_{i}(\tau)\left[\psi_{i}(\tau)+Ic_{i\downarrow}(\tau)c_{i\uparrow}(\tau)-\gamma_{i}(\tau)\right]+c.c.\right\} ,
\end{eqnarray}
we have
\begin{eqnarray}
Z\left[\bar{\gamma,}\gamma\right] & = & Z_{0}^{B}\det\left[\mathbf{J}^{-1}\right]\det\left[-\mathbf{J}\right]\int\mathcal{D}\left[\bar{c},c,\bar{\psi},\psi,\bar{\phi},\phi\right]e^{\sum_{ij}\int_{0}^{\hbar\beta}d\tau J_{ij}\bar{\phi}_{i}(\tau)\phi_{j}(\tau)+\sum_{i}\int_{0}^{\hbar\beta}d\tau\left\{ \bar{\phi}_{i}(\tau)\psi_{i}(\tau)+c.c.\right\} },\nonumber \\
 &  & \times e^{-\frac{1}{\hbar}W_{1}\left[\bar{\psi,}\psi\right]+\tilde{S}_{0}^{F}\left[\bar{c,}c,\bar{\Delta},\Delta\right]+\sum_{i}\int_{0}^{\hbar\beta}d\tau\left\{ \bar{\phi}_{i}(\tau)\gamma_{i}(\tau)+c.c.\right\} }.\label{eq:generarting-global-partition-function 2}
\end{eqnarray}
From Eqs. (\ref{eq:generarting-global-partition-function}) and (\ref{eq:generarting-global-partition-function 2}),
we see that the $b_{i}(\tau)$, $\bar{b}_{i}(\tau)$ and $\phi_{i}(\tau)$,
$\bar{\phi}_{i}(\tau)$ fields have the same generating functional $Z\left[\bar{\gamma,}\gamma\right]$. The above considerations about generating functional correspond to those in Appendix A of Ref. \cite{2005PhRvA..71c3629S}.
\end{widetext}

\subsection{Mean-field equations for order parameters - the operator approach
\label{sub: meanfield - lienar response}}

Eqs. (\ref{eq: equation coupled}) were derived by using coherent
state path integral within double Hubbard-Stratonovich transformation
within the bosonic part of action. Now, we show that these equations
can be also recovered by using a standard operator approach, at least
in the small $\phi_{0}$ limit. In order to get the equations for
order parameters $\phi_{0}$ and $x_{0}$, we start from the mean-field
approximation applied to the BFHM Hamiltonian defined in Eq. (\ref{eq:BFM-hamiltonian}),
i.e.

- for bosonic hopping term:
\begin{eqnarray}
-\sum_{ij}J_{ij}b_{i}^{\dagger}b_{j} & \approx & NzJ\left|\phi_{0}\right|^{2}-zJ\phi_{0}\sum_{i}b_{i}^{\dagger}-zJ\bar{\phi}_{0}\sum_{i}b_{i}\,,
\end{eqnarray}

- for fermionic interaction term (BCS type approximation in the pairing
channel): 
\begin{eqnarray}
 &  & V\sum_{i}c_{i\uparrow}^{\dagger}c_{i\downarrow}^{\dagger}c_{i\downarrow}c_{i\uparrow}\nonumber \\
 &  & \approx\frac{V}{N}\sum_{\mathbf{k}\mathbf{k}'}c_{\mathbf{k}'\uparrow}^{\dagger}c_{-\mathbf{k}'\downarrow}^{\dagger}c_{-\mathbf{k}\downarrow}c_{\mathbf{k}\uparrow}\nonumber \\
 &  & \approx-\frac{N}{V}\left|\Delta_{0}\right|^{2}+\sum_{\mathbf{k}}\bar{\Delta}_{0}c_{-\mathbf{k}\downarrow}c_{\mathbf{k}\uparrow}+\sum_{\mathbf{k}'}c_{\mathbf{k}'\uparrow}^{\dagger}c_{-\mathbf{k}'\downarrow}^{\dagger}\Delta_{0}\,,
\end{eqnarray}

- for resonant interaction term:
\begin{eqnarray}
 &  & I\sum_{i}\left(c_{i\uparrow}^{\dagger}c_{i\downarrow}^{\dagger}b_{i}+b_{i}^{\dagger}c_{i\downarrow}c_{i\uparrow}\right)\nonumber \\
 &  & \approx I\sum_{\mathbf{k}}\left(c_{\mathbf{k}\uparrow}^{\dagger}c_{-\mathbf{k}\downarrow}^{\dagger}\phi_{0}+\bar{\phi}_{0}c_{-\mathbf{k}\downarrow}c_{\mathbf{k}\uparrow}\right)\nonumber \\
 &  & +I\frac{1}{V}\sum_{i}\left(\bar{\Delta}_{0}b_{i}+\Delta_{0}b_{i}^{\dagger}\right)-I\frac{1}{V}\sum_{i}\left(\bar{\Delta}_{0}\phi_{0}+\Delta_{0}\bar{\phi}_{0}\right)\,.
\end{eqnarray}
Then, the thermodynamic potential can be written in the form

\begin{equation}
\Omega=-\frac{1}{\beta}\ln Z,\label{eq: thermodynamic potential}
\end{equation}
with
\[
Z=\textrm{Tr}\, e^{-\beta\left(H_{eff}^{F}+H_{eff}^{B}+H_{eff}^{FB}\right)}
\]
and where
\begin{eqnarray}
H_{eff}^{F} & = & \sum_{\mathbf{k}\sigma}\xi_{\mathbf{k}}c_{\mathbf{k}\sigma}^{\dagger}c_{\mathbf{k}\sigma}-\sum_{\mathbf{k}}\left(\bar{\Delta}_{0}-I\bar{\phi}_{0}\right)c_{-\mathbf{k}\downarrow}c_{\mathbf{k}\uparrow}\nonumber \\
 &  & -\sum_{\mathbf{k}}c_{\mathbf{k}\uparrow}^{\dagger}c_{-\mathbf{k}\downarrow}^{\dagger}\left(\Delta_{0}-I\phi_{0}\right)+\frac{N}{V}\left|\Delta_{0}\right|^{2},
\end{eqnarray}
\begin{eqnarray}
H_{eff}^{B} & = & NzJ\left|\phi_{0}\right|^{2}+\left(I\frac{1}{V}\Delta_{0}-zJ\phi_{0}\right)\sum_{i}b_{i}^{\dagger}\nonumber \\
 &  & +\left(I\frac{1}{V}\bar{\Delta}_{0}-zJ\bar{\phi}_{0}\right)\sum_{i}b_{i}-\sum_{i}\mu^{*}b_{i}^{\dagger}b_{i}\nonumber \\
 &  & +U\sum_{i}b_{i}^{\dagger}b_{i}^{\dagger}b_{i}b_{i}\,,
\end{eqnarray}
\begin{equation}
H_{eff}^{FB}=-I\frac{N}{V}\left(\bar{\Delta}_{0}\phi_{0}+\Delta_{0}\bar{\phi}_{0}\right).
\end{equation}
Next, the $\phi_{0}$ and $\Delta$ amplitudes can be obtained from
the conditions
\begin{equation}
\frac{\partial\Omega}{\partial\bar{\Delta}_{0}}=0,\ \ \ \ \ \ \frac{\partial\Omega}{\partial\bar{\phi}_{0}}=0,\label{eq: extremum conditions}
\end{equation}
which give
\begin{equation}
\left\{ \begin{array}{l}
0=-\frac{N}{V}\Delta_{0}+I\frac{N}{V}\phi_{0}+\sum_{\mathbf{k}}\left\langle c_{-\mathbf{k}\downarrow}c_{\mathbf{k}\uparrow}\right\rangle -\frac{I}{V}\sum_{i}\left\langle b_{i}\right\rangle ,\\
0=-I\sum_{\mathbf{k}}\left\langle c_{-\mathbf{k}\downarrow}c_{\mathbf{k}\uparrow}\right\rangle -NzJ\phi_{0}+zJ\sum_{i}\left\langle b_{i}\right\rangle +I\frac{N}{V}\Delta_{0}.
\end{array}\right.
\end{equation}
This leads to
\begin{equation}
x_{0}=\frac{1}{N}\sum_{\mathbf{k}}\left\langle c_{-\mathbf{k}\downarrow}c_{\mathbf{k}\uparrow}\right\rangle ,\label{eq: Delta}
\end{equation}
\begin{equation}
\phi_{0}=\frac{1}{N}\sum_{i}\left\langle b_{i}\right\rangle ,\label{eq: phi_0}
\end{equation}
where in this section statistical average is defined as $\left\langle ...\right\rangle =\textrm{Tr}\,...\, e^{-\beta\left(H_{eff}^{fer}+H_{eff}^{bos}+H_{eff}^{fer-bos}\right)}/Z$
and we introduce $x_{0}=\Delta/V$ the same as in Sec. \ref{sub:Saddle-point-of the effective action}.

Now we focus on the first equation, i.e. Eq. (\ref{eq: Delta}). Expectation
value $\left\langle c_{-\mathbf{k}\downarrow}c_{\mathbf{k}\uparrow}\right\rangle $
for a given wave vector $\mathbf{k}$ can be calculated by diagonalizing
$H_{eff}^{fer}$ Hamiltonian using the standard Bogoliubov transformation
\begin{equation}
c_{\mathbf{k}\uparrow}=\bar{u}_{\mathbf{k}}\gamma_{\mathbf{k}\uparrow}+\bar{v}_{\mathbf{k}}\gamma_{-\mathbf{k}\downarrow}^{\dagger}\,,
\end{equation}
\begin{equation}
c_{\mathbf{k}\downarrow}=\bar{u}_{\mathbf{k}}\gamma_{\mathbf{k}\downarrow}-\bar{v}_{\mathbf{k}}\gamma_{-\mathbf{k}\uparrow}^{\dagger}\,,
\end{equation}
with
\begin{equation}
\left|u_{\mathbf{k}}\right|^{2}=\frac{1}{2}\left(1+\frac{\xi_{\mathbf{k}}}{E_{\mathbf{k}}}\right),
\end{equation}
\begin{equation}
\left|v_{\mathbf{k}}\right|^{2}=\frac{1}{2}\left(1-\frac{\xi_{\mathbf{k}}}{E_{\mathbf{k}}}\right),
\end{equation}
then we obtain
\begin{equation}
\left\langle c_{-\mathbf{k}\downarrow}c_{\mathbf{k}\uparrow}\right\rangle =\frac{Vx_{0}-I\phi_{0}}{2E_{\mathbf{k}}}\tanh\left(\frac{\beta}{2}E_{\mathbf{k}}\right)\,,\label{eq: <cc>}
\end{equation}
with a quasi-particle fermionic energy $E_{\mathbf{k}}$ defined as
before in Eq. (\ref{eq: EkF}). 

Next equation, i.e. Eq. (\ref{eq: phi_0}), we calculate by using
the linear response theory. Assuming, that $\phi_{0}$ and $x_{0}$
amplitudes are small one can expand $\left\langle b_{i}\right\rangle $
in terms of these parameters which gives 
\begin{eqnarray}
\frac{1}{N}\sum_{i}\left\langle b_{i}\right\rangle  & \approx & -\frac{1}{\hbar}zJ\phi_{0}G^{1,c}\left(i\nu_{n}=0\right)\nonumber \\
 &  & +\frac{1}{\hbar}Ix_{0}G^{1,c}\left(i\nu_{n}=0\right)\,,\label{eq: <b>}
\end{eqnarray}
\begin{widetext}Finally, combining Eqs. (\ref{eq: Delta}, \ref{eq: phi_0},
\ref{eq: <cc>}, \ref{eq: <b>}), one gets
\begin{equation}
\left\{ \begin{array}{l}
\left(\epsilon_{\mathbf{0}}-\hbar\left[G^{1,c}\left(i\nu_{n}=0\right)\right]^{-1}\right)\phi_{0}=-\frac{I}{N}\sum_{\mathbf{k}}\frac{Vx_{0}-I\phi_{0}}{2E_{\mathbf{k}}^{F}}\tanh\left(\frac{\beta}{2}E_{\mathbf{k}}\right),\\
x_{0}=\frac{1}{N}\sum_{\mathbf{k}}\frac{Vx_{0}-I\phi_{0}}{2E_{\mathbf{k}}}\tanh\left(\frac{\beta}{2}E_{\mathbf{k}}\right),
\end{array}\right.\label{eq: result from operator}
\end{equation}
which recovers the result from coherent state path integral, i.e.
Eqs. (\ref{eq: equation coupled}) in the limit of small $\phi_{0}$,
in which the term $gN\hbar\beta\left|\phi_{0}\right|^{2}\phi_{0}$
can be neglected (i.e. on the phase boundary). 

Moreover, it is also worth adding that the above derivation of equations
for order parameters $x_{0}$ and $\phi_{0}$ (i.e. Eq. (\ref{eq: result from operator})),
can be also handled by using an explicit form of thermodynamic potential
\begin{equation}
\Omega=\Omega_{F}+\Omega_{FB}+\Omega_{B}\,,\label{eq: Omega}
\end{equation}
where
\begin{equation}
\Omega_{F}/N=\frac{1}{N}\sum_{\mathbf{k}}\left(\xi_{\mathbf{k}}-E_{\mathbf{k}}\right)+V\left|x_{0}\right|^{2}-\frac{2}{\beta N}\sum_{\mathbf{k}}\ln\left(1+e^{-\beta E_{\mathbf{k}}}\right),
\end{equation}
\begin{equation}
\Omega_{FB}/N=-I\left(\bar{x}_{0}\phi_{0}+x_{0}\bar{\phi}_{0}\right),
\end{equation}
\begin{eqnarray}
\Omega_{B} & /N= & -\frac{1}{\beta}\ln\textrm{Tr}e^{-\beta\left(zJ\left|\phi_{0}\right|^{2}+\left(Ix_{0}-zJ\phi_{0}\right)b_{i}^{\dagger}+\left(I\bar{x}_{0}-zJ\bar{\phi}_{0}\right)b_{i}-\mu^{*}b_{i}^{\dagger}b_{i}+Ub_{i}^{\dagger}b_{i}^{\dagger}b_{i}b_{i}\right)}.\label{eq: Omega-bosons}
\end{eqnarray}
Then extremizing $\Omega$ in terms of $\bar{x}_{0}$ and $\bar{\phi}_{0}$
yields general mean-field equations for order parameters
\begin{equation}
x_{0}=\frac{1}{N}\sum_{\mathbf{k}}\frac{Vx_{0}-I\phi_{0}}{2E_{\mathbf{k}}}\tanh\left(\frac{\beta}{2}E_{\mathbf{k}}\right)
\end{equation}
\begin{equation}
\phi_{0}=\frac{1}{N}\sum_{i}\left\langle b_{i}\right\rangle _{B}\label{eq: phi0}
\end{equation}
where $\left\langle ...\right\rangle _{B}=\textrm{Tr}\,...\, e^{-\beta H_{eff}^{bos}}/Z$,
$Z=\textrm{Tr}\, e^{-\beta H_{eff}^{bos}}$ and should be compared
to Eqs (\ref{eq: result from operator}) or (\ref{eq: equation coupled})
which was evolved close to the phase boundary. Moreover, from Eqs.
(\ref{eq: Omega}-\ref{eq: Omega-bosons}) it is easy to notice that
the thermodynamic potential $\Omega$ consists of standard BCS-like
part $\Omega_{fer}$, BHM-like part $\Omega_{bos}$ and part $\Omega_{fer-bos}$
which is proportional to Feshbach interaction energy $I$. Eqs. (\ref{eq: Omega}-\ref{eq: phi0})
make also a clear framework for further analysis of thermodynamic
properties of BFHM. As an example the free energy $F$ is now simply
given by $F/N=\Omega/N+\mu n$ in which 
\begin{equation}
n=-\frac{1}{N}\frac{\partial\Omega}{\partial\mu}=n_{F}+2n_{B}
\end{equation}
\begin{equation}
n_{F}=\frac{1}{N}\sum_{\mathbf{k}}\left[1-\frac{\xi_{\mathbf{k}}}{E_{\mathbf{k}}}\tanh\left(\frac{\beta}{2}E_{\mathbf{k}}\right)\right]
\end{equation}
\begin{equation}
n_{B}=\frac{1}{N}\sum_{i}\left\langle b_{i}^{\dagger}b_{i}\right\rangle _{B}
\end{equation}
These mean-field results should be also compared with Eqs. (\ref{eq:nFB}-\ref{eq:nB})
in which the $0th$ order approximation was imposed on statistical
sum. Interestingly, the form of $\Omega_{B}$ and $\phi_{0}$ given
in Eqs. (\ref{eq: Omega-bosons}) and (\ref{eq: phi0}) can be calculated
exactly for limiting cases of hard-core bosonic interaction ($U\rightarrow\infty$)
and for the case where $U$ vanishes ($U=0$). For example within
the hard-core limit on-site bosonic density basis is restricted to
two occupation numbers (i.e. to 0 or 1 boson per site) and then one
gets $\Omega_{bos}/N=zJ\left|\phi_{0}\right|^{2}-\mu^{*}-\ln\left[2\cosh\left(\beta E_{g}\right)\right]/\beta$
where $E_{g}=\sqrt{\left(\mu^{*}\right)^{2}+\left|Ix_{0}-zJ\phi_{0}\right|^{2}}$
and for order parameter $\phi_{0}$ one finds $\phi_{0}=-\left(Ix_{0}-zJ\phi_{0}\right)\tanh\left(\beta E_{g}\right)/2E_{g}$
\cite{PhysRevB.36.180}.

At the end of this section, we would like to also add that going beyond
the critical line toward SF phase, it is worth mentioning that the
functional integral approach presented in Sec. \ref{sec:Model} and
the operator approach discussed here give different descriptions.
Indeed, evaluation of the expansion in Eq. (\ref{eq: <b>}) to the
third order in the $\phi_{0}$ and $x_{0}$ amplitudes, generates
coefficients with four point local bosonic correlation function denoted
by $G_{i}^{2,c}(\tau{}_{1}',\,\tau'_{2},\,\tau{}_{1},\,\tau{}_{p})$
(see Eq. (\ref{eq: Gpc})), while the path integral method gives $\Gamma_{i}^{2,c}\left(\tau,\tau',\tau'',\tau'''\right)$
(see Eq. (\ref{eq: Gamma_pc})). This higher order term in the path
integral formulation is denoted by $g$ in Eq. (\ref{eq: equation coupled}),
which is proportional to $\Gamma_{i}^{2,c}$ in the static limit.
Therefore, on the grounds of the previous considerations within the
BHM in Ref. \cite{2005PhRvA..71c3629S} we would like to point out,
that our path integral formulation, should be more relevant than the
operator ones, because its gives better description of gaussian fluctuation
in the BHM limit with SF phase.

\end{widetext}

\bibliographystyle{apsrev}
\bibliography{library}

\end{document}